# Highly ordered lead-free double perovskite halides by design

Chang Won Ahn[a,†], Jae Hun Jo[a,†], Jong Chan Kim[b], Hamid Ullah[c], Sangkyun Ryu[d], Younghun Hwang[e], Jin San Choi[a], Jongmin Lee[f], Sanghan Lee[f], Hyoungjeen Jeen[d], Young-Han Shin[c], Hu Young Jeong[b], Ill Won Kim[a], and Tae Heon Kim[a]*

[a]*Department of Physics and Energy Harvest Storage Research Center (EHSRC), University of Ulsan, Ulsan 44610, Republic of Korea.*

[b]*UNIST Central Research Facilities (UCRF) & School of Materials Science and Engineering, Ulsan National Institute of Science and Technology (UNIST), Ulsan 44919, Republic of Korea.*

[c]*Multiscale Materials Modelling Laboratory, Department of Physics, University of Ulsan, Ulsan 44610, Republic of Korea.*

[d]*Department of Physics, Pusan National University, Pusan 46241, Republic of Korea.*

[e]*School of Electrical and Electronics Engineering, Ulsan College, Ulsan 44610, Republic of Korea.*

[f]*School of Materials Science and Engineering, Gwangju Institute of Science and Technology, Gwangju 61005, Republic of Korea.*

[*]Electronic mail: thkim79@ulsan.ac.kr

[†]These authors contributed equally to this work.

**Abstract**

Lead-free double perovskite halides are emerging optoelectronic materials that are alternatives to lead-based perovskite halides. Recently, single-crystalline double perovskite halides were synthesized, and their intriguing functional properties were demonstrated. Despite such pioneering works, lead-free double perovskite halides with better crystallinity are still in demand for applications to novel optoelectronic devices. Here, we realized highly crystalline $Cs_2AgBiBr_6$ single crystals with a well-defined atomic ordering on the microscopic scale. We avoided the formation of Ag vacancies and the subsequent secondary $Cs_3Bi_2Br_9$ by manipulating the initial chemical environments in hydrothermal synthesis. The suppression of Ag vacancies allows us to reduce the trap density in the as-grown crystals and to enhance the carrier mobility further. Our design strategy is applicable for fabricating other lead-free halide materials with high crystallinity.

1. **Introduction**

For the last few decades, lead-based perovskite halides CsPb$X_3$ ($X$ = Cl, Br, I) have been promising candidates in the field of optoelectronic devices, which include photovoltaic solar cells,[1-3] X-ray detectors,[4-6] and light emitting diodes.[7-9] Despite their excellent functionality (*e.g.*, the high solar-cell efficiency of 23.7%),[10] their use in actual devices has been limited due to the global regulation of toxic lead.[11] Very recently, alternative materials that do not contain a Pb$^{2+}$ ion at the perovskite B-site and are eco-friendly have been under intensive exploration.[12-15]

Currently, lead-free double perovskite halides Cs$_2$AgBi$X_6$ are of enormous interest as alternatives to lead-based perovskite halides CsPb$X_3$.[9,16,17] Here, two neighbouring Pb$^{2+}$ ions in the perovskite halides are alternately replaced with monovalent Ag$^+$ and trivalent Bi$^{3+}$ ions resulting in a double perovskite structure, as shown in Figure 1a. Some pioneers have already synthesized lead-free double perovskite halide single crystals with good crystallinity. These materials showed fascinating physical properties, including a long lifetime in carrier recombination,[18] low effective mass,[18] robust phase stability against humidity,[18,19] and white-light emission by self-trapped excitons.[20] Nevertheless, Cs$_2$AgBi$X_6$ single crystals with better crystallinity are still in demand for potential applications in a wide range of optoelectronic devices.[21] To achieve this, a systematic study of the synthesis of highly crystalline Cs$_2$AgBi$X_6$ single crystals is essential.

The structural stability of lead-free double perovskite halide Cs$_2$AgBi$X_6$ is very susceptible to changes in the ambient chemical composition during single crystal growth. It has been theoretically demonstrated that various defects in a double perovskite Cs$_2$AgBiBr$_6$ can be created, such as Ag vacancies, Bi vacancies, and Ag$_{Bi}$ anti-site defects.[22] Note that their formation energies in theoretical calculations are dependent on a particular chemical condition (*e.g.*, Ag-rich and Bi-rich). It is also interesting that the formation energies of secondary phases, which include CsAgBr$_2$ (tetragonal, *P4/nmm*),[23] Cs$_2$AgBr$_3$ (orthorhombic, *Pnma*),[23] and Cs$_3$Bi$_2$Br$_9$ (trigonal, *P-3m1*),[24] compete with that of Cs$_2$AgBiBr$_6$ in the thermodynamic reaction process. Furthermore, it appears that the most energetically stable phase is very different relying on the given chemical environment. Despite these intriguing thermodynamic calculations of the phase stability in lead-free double perovskite halides, it has rarely been examined how the structural phase of $Cs_2AgBiX_6$ during crystal growth evolves depending on variations in either the $Ag^+$ or $Bi^{3+}$ content.

It is worth noting that the chemical stoichiometry and crystallinity of $Cs_2AgBiX_6$ single crystals can strongly depend on the compositional ratio between $Ag^+$ and $Bi^{3+}$ ions during crystal growth. Due to the high ionic conductivity of $Ag^+$ ions,[25-27] some of the mobile $Ag^+$ ions would not participate in the crystallization, and then the rate of chemical reaction can be dependent on the total amount of $Ag^+$ cations inside a precursor solution. Although a stoichiometric amount of $Ag^+$ ions are incorporated in the precursor solution, the resulting $Cs_2AgBiX_6$ crystal will be Ag-deficient (*i.e.*, Bi-excessive) due to the itinerant $Ag^+$ ions. This results in poor crystallinity with the appearance of impurity/secondary phases. On the other hand, it is well known that a transition-metal Bi atom is very volatile, and thus it can be easily vaporized in a thermally-assisted chemical reaction process.[28-30] To avoid Bi deficiency in the end product, a high content of $Bi^{3+}$ ions exceeding the stoichiometric composition was added in the beginning stage of material synthesis.[28-30] Therefore, it is of great interest to investigate the effect of the initial $Ag^+$ or $Bi^{3+}$ content on the chemical composition and structural phases of the as-grown $Cs_2AgBiX_6$ single crystals.

In this work, we experimentally demonstrated the impact of Ag-excess or Bi-excess on the single crystal growth of a lead-free double perovskite halide $Cs_2AgBiBr_6$. While tuning the molar concentration of the reactants (*i.e.*, CsBr, AgBr and $BiBr_3$) in the hydrothermal synthesis of $Cs_2AgBiBr_6$, we systematically monitored how its crystal growth evolved as a function of the initial contents of the Ag and Bi elements in the AgBr and $BiBr_3$, respectively. Note that an Ag-rich (Bi-rich) environment is attainable in the as-prepared precursor solution for the hydrothermal reaction, as the concentration of the starting material AgBr ($BiBr_3$) exceeds the stoichiometric amount. We found that the degree of crystallization strongly depends on the initial reaction conditions. It appeared that the as-grown $Cs_2AgBiBr_6$ single crystals under Ag-rich conditions are highly crystalline with a well-defined double perovskite structure microscopically. In contrast, under Bi-rich conditions, the crystallization of $Cs_2AgBiBr_6$ is relatively poor with the formation of a parasitic $Cs_3Bi_2Br_9$ phase. We also identified that the Cs$_2$AgBiBr$_6$ single crystals synthesized under Ag-excess conditions exhibited a larger indirect band gap ($E_{g,indirect}$ ~ 2.12 eV) than those ($E_{g,indirect}$ ~ 2.07 eV) synthesized under the stoichiometric conditions, which is close to the value predicted ($E_{g,indirect}$ ~ 2.26 eV) by theoretical calculations. A possible origin of such a difference in the crystallinity of Cs$_2$AgBiBr$_6$ will be discussed in conjunction with its effect on the electrical transport properties (*e.g.*, carrier mobility) of the as-grown single crystals.

To fabricate lead-free double perovskite halide Cs$_2$AgBiBr$_6$ single crystals, we used a conventional hydrothermal reaction technique. In the hydrothermal method, a reactant solution in a Teflon vessel was pressurized with an autoclave during the crystal growth. Under the internal pressure imposed by the autoclave, a double perovskite Cs$_2$AgBiBr$_6$ phase was easily stabilized, enabling the subsequent growth of single crystals. For the hydrothermal reaction, we first prepared a starting solution by dissolving high-purity CsBr (99.9%), AgBr (99.0%), and BiBr$_3$ (≥ 98.0%) powders in a buffered HBr solvent (the HBr weight percentage of 48% in H$_2$O).[18] As shown in Figure 1b, the as-prepared precursor solution was fired in a box furnace up to 130 ˚C with a ramping rate of 5 ˚C/min. Then, it was cooled down to room temperature with a ramping rate of 1 ˚C/hour for crystallization. For more details related to the single crystal growth, see the method section and Figure S1 (Supporting Information).

In the hydrothermal synthesis of Cs$_2$AgBiBr$_6$ single crystals, it is possible to control the initial chemical environment of the reaction. A chemical reaction between the starting substances (*i.e.*, CsBr, AgBr, and BiBr$_3$) produces a stoichiometric Cs$_2$AgBiBr$_6$ as follows:[22]

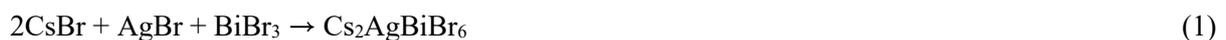

2CsBr + AgBr + BiBr$_3$ → Cs$_2$AgBiBr$_6$            (1)

Note that the ratio of molar concentrations between AgBr and BiBr$_3$ is 1 in the stoichiometric case. We stress that it is important to assess the degree of crystallization of double perovskite halides depending on the initial concentrations of Ag$^+$ and Bi$^{3+}$ ions that take part in a chemical reaction. Thus, we produced an excess Ag or excess Bi condition while varying the molar concentrations of the AgBr and BiBr$_3$ reactants in a precursor solution as follows (Figure 1c):

$$2CsBr_{(s)} + (1+x)\ AgBr_{(s)} + (1+y)\ BiBr_{3(s)} + HBr_{(aq)} \rightarrow Cs_2AgBiBr_{6(s)} + (x\ Ag^{1+} + y\ Bi^{3+} + HBr)_{(aq)} \quad (2)$$

Herein, the $x$ ($y$) value was varied from 0.05 to 0.20 with increments of 0.05 in the Ag (Bi)-excess condition, whereas the $y$ ($x$) value was fixed to zero, as shown Figure 1d. For example, when the $x$ and $y$ values were 0.05 and 0.00, respectively, the corresponding molar concentrations of $(1+x)$ AgBr and $(1+y)$ BiBr$_3$ reactants became 1.05 and 1.00, respectively, which allowed us to artificially generate the Ag-excess condition in the hydrothermal reaction. In the opposite case (*i.e.*, $x = 0$ and $y > 0$), the Bi-excess condition was attainable as well.

We found that the lateral sizes of Cs$_2$AgBiBr$_6$ single crystals grown by a hydrothermal method strongly depended on the initial molar concentrations of AgBr and BiBr$_3$ reactants (Figure 1e). More details of the size estimation of the as-grown Cs$_2$AgBiBr$_6$ single crystals are described in Figure S2 (Supporting Information). It is also interesting that the double perovskite Cs$_2$AgBiBr$_6$ phase is highly crystallized with an average lateral dimension of ~4 mm under an excess Ag environment only, although there were some variations in the measured crystal size (the measured values of the lateral sizes of all Cs$_2$AgBiBr$_6$ crystals used in this work are shown in Table S1). In contrast, under excess Bi conditions, the crystallinity of the as-produced Cs$_2$AgBiBr$_6$ crystals was very poor and still remained in a powder form. For the stoichiometric case, where the molar ratio between Ag$^+$ and Bi$^{3+}$ contents was 1, the measured crystal size quite varied significantly from ~1 to ~5 mm, indicating that it was difficult to achieve reproducible growth of Cs$_2$AgBiBr$_6$ single crystals with high crystallinity.

## 2. Results and Discussion

Figure 2a shows the powder x-ray diffraction (XRD) results of Cs$_2$AgBiBr$_6$ single crystals grown under various chemical environments (*i.e.*, stoichiometric, excess Bi, and excess Ag conditions). To perform these measurements, Cs$_2$AgBiBr$_6$ single crystals were first synthesized using a precursor solution with different molar concentrations of Ag$^+$ [$(1+x)$] and Bi$^{3+}$ [$(1+y)$] ions by a hydrothermal method; then, we prepared $Cs_2AgBiBr_6$ powders by grinding the as-grown $Cs_2AgBiBr_6$ single crystals (for more details of the synthesis of $Cs_2AgBiBr_6$ single crystals, see the method section in Supporting Information). We found that the obtained powder XRD patterns (the blue solid lines) of $Cs_2AgBiBr_6$ single crystals grown under Ag-excess conditions exactly matched that of a cubic $Cs_2AgBiBr_6$ phase ($a = b = c = 11.25$ Å) (detailed analyses of powder XRD results, Figure S3, Supporting Information).[18] In contrast, it appeared that the as-grown $Cs_2AgBiBr_6$ single crystals under Bi-excess conditions contained a secondary $Cs_3Bi_2Br_9$ phase (marked by the red solid circles) and a little amount of $BiBr_3$ residues (Figure S3b, Supporting Information) in addition to the major double perovskite phase (marked by the blue solid diamonds) (Figure 2b). For comparison, the measured XRD pattern of $Cs_3Bi_2Br_9$ powders is shown in the lowest panel of Figure 2a (a detailed comparison of the obtained XRD patterns of $Cs_2AgBiBr_6$ powders with the reference XRD pattern of $Cs_3Bi_2Br_9$ is provided in Figure S4, Supporting Information).

Figure 2c shows the volume fractions of the primary $Cs_2AgBiBr_6$ and secondary $Cs_3Bi_2Br_9$ phases. We estimated the relative portions of $Cs_2AgBiBr_6$ and $Cs_3Bi_2Br_9$ phases by fitting the (111) and ($10\bar{1}0$) Bragg peaks (around the $2\theta$ angle of 13°) of double perovskite $Cs_2AgBiBr_6$ and layered perovskite $Cs_3Bi_2Br_9$ phases using a Lorentzian distribution function. It is evident that the volume fraction of the primary $Cs_2AgBiBr_6$ (the secondary $Cs_3Bi_2Br_9$) phase keeps increasing (decreasing) as the molar concentration of $Bi^{3+}$ ions decreases from 1.2 to 1.0 under excess Bi conditions. Here, the molar content of $Ag^+$ ions relative to the Bi content increased, although the absolute molar concentration of $Ag^+$ ions was fixed at 1. It is also noticeable that the secondary $Cs_3Bi_2Br_9$ phase (the extracted volume fraction of ~4%) still remains in $Cs_2AgBiBr_6$ single crystals synthesized under stoichiometric conditions (denoted by a black arrow). In contrast, there is no $Cs_3Bi_2Br_9$ phase in the as-grown single crystals under excess Ag conditions, and the $Cs_2AgBiBr_6$ phase becomes dominant. This indicates that the phase stability of double perovskite $Cs_2AgBiBr_6$ is very susceptible to the initial chemical concentrations in the precursor solution prepared for the hydrothermal reaction.

To gain further insight into the chemical-environment-dependent structural instability in $Cs_2AgBiBr_6$ single crystals, we carried out energy-dispersive x-ray spectroscopy (EDX) measurements to analyze the chemical stoichiometry. As shown in Figure 3a, the element-specific EDX results reveal that the Ag and Bi contents in the as-synthesized $Cs_2AgBiBr_6$ compounds are very different depending on the initial molar concentrations of $Ag^+$ and $Bi^{3+}$ ions in a precursor solution. On the other hand, there was no significant change in the measured Cs and Br contents, which are almost constant with atomic percentages of 20 and 60%, respectively [For the stoichiometric analyses of as-synthesized $Cs_2AgBiBr_6$ single-crystal/powder compounds, we carrier out scanning electron microscopy (SEM) and EDX measurements. And, the obtained SEM images and EDX spectra are shown in Figure 3b-i]. Considering the fact that the atomic percentages of Cs, Ag, Bi, and Br elements are 20, 10, 10, and 60% in stoichiometric $Cs_2AgBiBr_6$, respectively, all the $Cs_2AgBiBr_6$ single-crystal/powder specimens should contain stoichiometric amount of $Cs^+$ and $Br^-$ ions.

To further assess either Bi or Ag deficiencies in the $Cs_2AgBiBr_6$ compounds, we also calculated the proportion of Bi to Ag contents from the measured atomic percentages of Bi and Ag elements (Figure 3j). For $Cs_2AgBiBr_6$ single crystals grown under excess Ag conditions, the atomic percentages of Ag and Bi elements are the same (approximately 10%), and the derived Bi/Ag values are close to 1, indicating that there are no Bi and Ag vacancies inside the as-grown single crystals. In contrast, it is clear that the Bi and Ag ratio is larger than 1 for $Cs_2AgBiBr_6$ compounds synthesized under both excess Bi and stoichiometric (denoted by a black arrow) conditions. This indicates that the measured single crystal/powder specimens have excessive Bi (*i.e.*, Ag-deficient) elements due to the presence of the secondary $Cs_3Bi_2Br_9$ phase.

High-resolution XRD measurements were performed to macroscopically examine the crystallinity of our $Cs_2AgBiBr_6$ single crystals (for details related to the XRD experiments, see the method section in Supporting Information). For these XRD analyses, we first selected a $Cs_2AgBiBr_6$ single crystal synthesized under an excess Ag condition where the initial molar concentrations of $Ag^+$ and $Bi^{3+}$ ions were 1.2 and 1.0, respectively. Note that $Cs_2AgBiBr_6$ single crystals grown under excess Ag conditions were well crystallized with no impurity phases, whereas the as-grown single crystals under excess Bi and stoichiometric conditions poorly crystallized and showed a secondary $Cs_3Bi_2Br_9$ phase (Figure 2a). The subsequent XRD $\theta$-$2\theta$ result clearly shows that the as-grown $Cs_2AgBiBr_6$ single crystals in an excess Ag environment were single-crystalline with a [111] crystallographic orientation (Figure 4a). An XRD phi ($\phi$) scan of the (220) Bragg peak shows 3-fold symmetry due to the preferred [111] orientation in a cubic double perovskite structure (Figure 4b). The as-grown single crystal should be in a single domain state structurally, because no diffraction peak due to in-plane misorientation was observed except for the three {220} Bragg peaks. To evaluate the mosaicity of our $Cs_2AgBiBr_6$ single crystals, we also carried out XRD rocking-curve measurements of the (111) Bragg peak, as shown in Figure 4c. The measured diffraction peak is very sharp, and then the estimated full width at half maximum (FWHM) *via* the best fit was 0.03°, which is smaller than the FWHM value of previously reported $Cs_2AgBiBr_6$ (0.06°) single crystal,[31] lead-based perovskite halide $CsPbBr_3$ (0.16°),[32] $CH_3NH_3PbBr_3$ (0.07°). This peak is comparable to conventional perovskite oxide $LiNbO_3$ (0.02°)[33] and $SrTiO_3$ (0.01°) single crystals (XRD rocking-curve results of $CH_3NH_3PbBr_3$ and $SrTiO_3$ single crystals are shown in Figure S5, Supporting Information). Accordingly, it is highly likely that there is no mosaic spread in our $Cs_2AgBiBr_6$ single crystals, and they should be highly crystallized with a single domain configuration.

To visualize the atomic structure in our $Cs_2AgBiBr_6$ single crystals, we carried out cross-sectional scanning transmission electron microscopy (STEM) experiments (sample preparation and STEM measurement details are provided in the method section and Figure S6, Supporting Information). As with the XRD analyses, we used a high crystalline $Cs_2AgBiBr_6$ single crystal synthesized under excess Ag conditions for the STEM measurements (Figure 5a). In double perovskite $Cs_2AgBiBr_6$, two neighboring halogen octahedra [*i.e.*, $AgBr_6$ (light cyan diamonds) and $BiBr_6$ (light yellow diamonds)] are alternately interconnected *via* corner sharing, resulting in an octahedral breathing order, as shown in Figure 5b. A high-angle annular dark field (HAADF)-STEM image of the as-grown $Cs_2AgBiBr_6$ single crystal revealed that the observed atomic configuration (Figure 5c) is in good agreement with the projected lattice structure (Figure 5b) along the [110] zone axis. It is worth noting that all the chemical elements (Cs, Ag, Bi, and Br) constituting $Cs_2AgBiBr_6$ single crystals were uniformly distributed throughout the whole region of the specimen with no spatial inhomogeneity (details about the STEM-EDX measurements are provided in Figure S7, Supporting

Information). In a fast Fourier transform (FFT) pattern obtained from the HAADF-STEM image, we observed {111} diffraction peaks arising from the lattice doubling of simple perovskite unit cells (Figure 5d). We also found that the FFT pattern corresponds to the simulated electron diffraction pattern of a cubic double perovskite structure (Figure 5e) (a comparison between the simulated electron diffraction patterns of cubic perovskite and double perovskite structures is shown in Figure S8, Supporting Information).

To identify the atomic arrangement of the as-grown $Cs_2AgBiBr_6$ single crystals, we plotted the line profiles of the measured STEM intensity in a HAADF image (Figure 5c) along the cubic [001] direction, as shown in Figure 5f. Note that a $Cs_2AgBiBr_6$ unit cell can be conceptually viewed as a stack of CsBr-BiBr$_2$ (or AgBr$_2$)-CsBr-AgBr$_2$ (or BiBr$_2$) in the [001] direction. Considering the fact that the peak intensities of atoms in the extracted line profiles are determined by the atomic number (Z),[34] the atomic stacking sequences in columns 1 (the red dashed box in Figure 5c) and 5 (the blue dashed box in Figure 5c) should correspond to Cs-Bi-Cs-Ag. And, in column 3 (the yellow dashed box in Figure 5c), the corresponding stacking sequence becomes Cs-Ag-Cs-Bi. When the $Cs_2AgBiBr_6$ unit cell is projected along the [110] zone axis, halogen Br atoms only appear with an interatomic spacing of 11.25 Å/2 (*i.e.*, a half lattice parameter of cubic $Cs_2AgBiBr_6$),[18] as displayed in columns 2 (the orange dashed box in Figure 5c) and 4 (the green dashed box in Figure 5c). Furthermore, in the intensity profiles, the peak position of each atom is very periodic, and there was no variation in the peak intensity. It follows that all the constituent atoms in our $Cs_2AgBiBr_6$ single crystals are highly ordered at the atomic level, resulting in the well-defined double perovskite structure.

Figure 6 shows space-charge-limited bulk conduction (SCLC) behaviors of $Cs_2AgBiBr_6$ single crystals synthesized under stoichiometric [(Ag$^+$, Bi$^{3+}$) = (1, 1) (Figure 6a)] and Ag-excess [(Ag$^+$, Bi$^{3+}$) = (1.05, 1), (1.1, 1), (1.15, 1), and (1.2, 1) (Figure 6b-e)] conditions, respectively. Note that the majority of the conduction carriers are holes (*i.e.*, p-type) in $Cs_2AgBiBr_6$, because the valence band maximum is close to the Fermi level (the calculated electronic band structure of $Cs_2AgBiBr_6$ is shown in Figure S9, Supporting Information).[31,35] A metal-insulator-metal (MIM) structure of Au/$Cs_2AgBiBr_6$/Au is commonly used for the hole-governing electrical transport measurements (for more details on our current density-voltage (*J-V*) measurements, see the method section and Figure S10, Supporting Information).[31,36] In the conventional SCLC model,[37-40] the *J-V* curves at low voltages (*i.e.*, $V < V_{tr}$) are characterized by linear ohmic behaviors ($J_{Ohm} \propto V$). As the voltage increases across a transition voltage of $V_{tr}$, the deep traps begin to fill, resulting in trap-filled-limited currents ($J_{TFL} \propto V^n$ with $n > 2$). At even higher voltages (*i.e.*, $V > V_{TFL}$), the deep traps are almost totally filled, exhibiting deep-trap-free-conduction behavior (*i.e.*, space-charge-limited currents, $J_{SCL} \propto V^2$). In this space-charge-limited region, the dark current density ($J_D$ fitted by the Mott-Gurney law and $V_{TFL}$)[37-40] is described as follows:

$$J_D = \frac{9\varepsilon\varepsilon_0\mu V^2}{8L^3} \qquad (3),$$

$$V_{TFL} = \frac{eN_tL^2}{2\varepsilon\varepsilon_0} \qquad (4),$$

where *e*, *ε*, *μ*, *L*, *V* are the charge of free carriers (*i.e.*, holes), dielectric constant, carrier mobility, sample thickness, and the applied voltage bias, respectively. $N_t$ is the density of deep traps. It is therefore possible for us to estimate the carrier mobility (*μ*) and trap density ($N_t$) by fitting the measured $J_{SCL}$-*V* curves with these formulas shown above.

The high crystallinity of our $Cs_2AgBiBr_6$ single crystals allowed us to achieve higher carrier mobility and lower trap density based on their *J-V* characteristics. To estimate the carrier mobility and trap density, we first measured the dielectric constants (*ε*) of the as-grown $Cs_2AgBiBr_6$ single crystals electrically (detailed information related to the dielectric permittivity measurements are in the method section and Figure S11, Supporting Information). Then, the *μ* and $N_t$ values of the as-grown $Cs_2AgBiBr_6$ single crystals were obtained *via* linear fits of the log *J*–log *V* plots). It is evident that the as-grown $Cs_2AgBiBr_6$ single crystals (*μ* and $N_t$ = 22.3 cm$^2$ V$^{-1}$ s$^{-1}$ and 9.57 × 10$^9$ cm$^{-3}$, respectively) under Ag-excess conditions (Figure 6e) exhibit ~5 times higher mobility and possess approximately three times lower trap density than those (*μ* and $N_t$ = 4.47 cm$^2$ V$^{-1}$ s$^{-1}$ and 36.0 × 10$^9$ cm$^{-3}$, respectively) under stoichiometric conditions (Figure 6a). Note that there was no noticeable difference in the surface morphologies of the as-synthesized $Cs_2AgBiBr_6$ single crystals under stoichiometric and Ag-excessive environments (Figure S12, Supporting Information), which indicates that a difference in the measured electrical properties was not attributed to extrinsic effects such as grain boundaries. More interestingly, the obtained $N_t$ decreased significantly as the initial $Ag^+$ molar concentration with respect to the $Bi^{3+}$ molar concentration increased (Figure S13a, Supporting Information). On the other hand, the estimated $\mu$ progressively increased with an increase in $Ag^+/Bi^{3+}$ molar ratio (Figure S13b, Supporting Information). The highest $\mu$ value (~22.3 cm$^2$ V$^{-1}$ s$^{-1}$) was measured in our $Cs_2AgBiBr_6$ single crystals synthesized under an Ag-abundant ($Ag^+$ : $Bi^{3+}$ = 1.2 : 1) environment. This value is comparable to the $\mu$ values (11.8 and 55.7 cm$^2$ V$^{-1}$ s$^{-1}$) reported in previous studies.[31,41]

To understand the microscopic origin of the synthetic-environment-dependent transport properties (*i.e.*, $\mu$ and $N_t$) in the as-grown $Cs_2AgBiBr_6$ single crystals, we predicted the formation energies of Ag and Bi vacancy defects through first-principles calculations, as shown in Table 1. To compute the defect formation energy ($E_{vf}$), we first added the energies of defective $Cs_2AgBiBr_6$ ($E_{total}$) (with either the Ag or Bi vacancy) and a single Ag or Bi atom ($E_{Ag/Bi}$). Then, the energy of defect-free $Cs_2AgBiBr_6$ ($E_0$) was subtracted from the sum of $E_{total}$ and $E_{Ag/Bi}$ (For more details on our theoretical calculations, see the method section in Supporting Information). It is interesting that the formation energy of an Ag vacancy (3.13 eV) is about twice as low as that of a Bi vacancy (6.15 eV). Thus, the formation of these Ag vacancy defects during crystal growth can be much easier under Ag-deficient and stoichiometric environments rather than Ag-excess conditions. At a given Ag vacancy site, the primary $Cs_2AgBiBr_6$ phase would be structurally unstable resulting in the emergence of a secondary $Cs_3Bi_2Br_9$ phase (a possible scenario of a structural transition from a double perovskite $Cs_2AgBiBr_6$ phase to a layered perovskite $Cs_3Bi_2Br_9$ phase is schematically described in Figure S14, Supporting Information). The UV-visible absorption measurements of our $Cs_2AgBiBr_6$ single crystals revealed that the as-grown single crystals (2.10 - 2.12 eV) under Ag-excess conditions exhibited a higher indirect band gap ($E_{g,indirect}$) than those (2.07 eV) under stoichiometric conditions (Figure S15 and Table S2, Supporting Information). Note that $E_{g,indirect}$ in $Cs_2AgBiBr_6$ was calculated to be ~2.26 eV in our theoretical results (Figure S9, Supporting Information). Considering the fact that an optical band gap in a solid is usually reduced with the formation of defect levels,[35,41] the as-grown $Cs_2AgBiBr_6$ single crystals under Ag-rich conditions should contain fewer Ag vacancies than those under stoichiometric conditions. Since vacancy defects in solids also act as trapping sites, impeding the movements in free charge carriers,[35,41] it is highly likely that the major hole carriers are more mobile in $Cs_2AgBiBr_6$ single crystals synthesized under an Ag excess environment due to the lower concentration of charge trapping sites, which is consistent with our transport results of $\mu$ and $N_t$.

To get further insight on the underlying mechanism of the enhanced transport properties in the as-grown $Cs_2AgBiBr_6$ single crystals under Ag-excessive conditions, we performed the temperature ($T$)-dependent conductivity measurements our $Cs_2AgBiBr_6$ single crystals, as shown in Figure 7. We first measured the current ($I$)-voltage ($V$) characteristics of two as-grown $Cs_2AgBiBr_6$ single crystals [*i.e.*, synthesized under stoichiometric (Figure 7a) and Ag-excessive (Figure 7b) conditions, respectively] in the temperature range from 298 to 373 K. Then, we plotted the measured electrical conductivity ($\sigma$) in accordance with the Arrhenius relation [$\ln(\sigma T)$ vs. $1/T$] and extracted the activation energy ($E_a$) for electrical transport *via* the subsequent linear fit of the Arrhenius plot (Figure 7c) [31,42]. For the $Cs_2AgBiBr_6$ single crystal under a stochiometric (Ag-excessive) condition, $E_a$ for charge transport was estimated to 0.80 (0.80) eV at high temperatures, whereas it was 0.37 (0.59) eV at low temperatures. Note that the high-temperature transport behaviors are dominated by thermal activation of major charge carriers (*i.e.*, mobile holes for $Cs_2AgBiBr_6$) and hence, the energy barrier height should be comparable to a half of an electronic band gap (*i.e.*, $E_{g,indirect}$ ~2.10 eV in our UV-visible absorption measurements of $Cs_2AgBiBr_6$ single crystals) in intrinsic semiconductors with slight variations [42]. On the contrary, $E_a$ at low temperatures can be easily affected by extrinsic effects such as vacancy defects and interfacial Schottky barrier [31,43]. Considering that the $Cs_2AgBiBr_6$ single crystal grown under a stoichiometric condition is more defective than the as-grown single crystal under an Ag-excessive condition, the charge transport at low temperatures would be governed by extrinsic defects with a lower $E_a$ than intrinsic charge carriers. Further studies are highly desirable to elucidate the mechanism of the defect-mediated electrical charge transport on an atomic scale.

## 3. Conclusions

In summary, we experimentally achieved highly crystalline lead-free double perovskite halide $Cs_2AgBiBr_6$ single crystals with a well-defined atomic ordering using a hydrothermal method. The successful growth of $Cs_2AgBiBr_6$ single crystals was achieved by systematically controlling the initial chemical environment in the hydrothermal synthesis. At the beginning stage of the hydrothermal reaction, Ag-rich conditions were used to suppress the formation of Ag vacancies in the single-crystal growth of $Cs_2AgBiBr_6$. Our results are of practical interest for fabricating high-quality lead-free halide materials and for optimizing their synthetic conditions where the reproducible growth of the halide materials is feasible. Conceptually, this work can be utilized to realize defect-free optoelectronic devices with high performance and multi-functionality.

**Acknowledgments**

T.H.K. acknowledges the National Research Foundation of Korea (NRF) grants funded by the Korea government (Ministry of Education) (NRF-2017R1D1A1B03028614 and NRF-2019R1A6A1A11053838). C.W.A acknowledges the support by Basic Science Research Program through the NRF funded the Ministry of Science and ICT (NRF-2018R1A2B6009210). Y.H.H. acknowledges the support by the NRF of Korea (NRF-2019R1I1A3A01063856). H.Y.J. acknowledges the support from Creative Materials Discovery Program (NRF-2016M3D1A1900035). Y.-H.S. acknowledges the support by Basic Science Research Program through the NRF funded the Ministry of Science and ICT (NRF-2018R1A2B6005159). Experiments at PLS-II were supported in part by MSICT and POSTECH.

**References**

[1] Lee MM, Teuscher J, Miyasaka T, Murakami TN, Snaith HJ. Efficient hybrid solar cells based on meso-superstructured organometal halide perovskites. Science 2012;338:643-647.

[2] Noh JH, Im SH, Heo JH, Mandal TN, Seok SI. Chemical management for colorful, efficient, and stable inorganic–organic hybrid nanostructured solar cells. Nano Lett 2013;13:1764-1769.

[3] Kojima A, Teshima K, Shirai Y, Miyasaka T. Organometal halide perovskites as visible-light sensitizers for photovoltaic cells. J Am Chem Soc 2009;131:6050-6051.

[4] Yakunin S, Sytnyk M, Kriegner D, Shrestha S, Richter M, Matt GJ, Azimi H, Brabec CJ, Stangl J, Kovalenko MV. Detection of x-ray photons by solutionprocessed lead halide perovskites. Nat Photonics 2015;9:444-449.

[5] Wei H, Fang Y, Mulligan P, Chuirazzi W, Fang H-H, Wang C, Ecker BR, Gao Y, Loi MA, Cao L. Sensitive x-ray detectors made of methylammonium lead tribromide perovskite single crystals. Nat Photonics 2016;10:333-339.

[6] Stoumpos CC, Malliakas CD, Peters JA, Liu Z, Sebastian M, Im J, Chasapis TC, Wibowo AC, Chung DY, Freeman AJ. Crystal growth of the perovskite semiconductor $CsPbBr_3$: a new material for high-energy radiation detection. Cryst Growth Des 2013;13:2722-2727.

[7] Tan Z-K, Moghaddam RS, Lai ML, Docampo P, Higler R, Deschler F, Price M, Sadhanala A, Pazos LM, Credgington D. Bright light-emitting diodes based on organometal halide perovskite. Nat Nanotechnol 2014;9:687-692.

[8] Cho H, Jeong S-H, Park M-H, Kim Y-H, Wolf C, Lee C-L, Heo JH, Sadhanala A, Myoung N, Yoo S. Overcoming the electroluminescence efficiency limitations of perovskite light-emitting diodes. Science 2015;350:1222-1225.

[9] Song J, Li J, Li X, Xu L, Dong Y, Zeng H. Quantum dot light-emitting diodes based on inorganic perovskite cesium lead halides ($CsPbX_3$). Adv Mater 2015;27:7162-7167.

[10] Min H, Kim M, Lee S-U, Kim H, Kim G, Choi K, Lee JH, Seok SI. Efficient, stable solar cells by using inherent bandgap of α-phase formamidinium lead iodide. Science 2019;366:749-753.

[11] Babayigit A, Ethirajan A, Muller M, Conings B. Toxicity of organometal halide perovskite solar cells. Nat Mater 2016;15:247-251.

[12] Saparov B, Hong F, Sun J-P, Duan H-S, Meng W, Cameron S, Hill IG, Yan Y, Mitzi DB. Thin-film preparation and characterization of $Cs_3Sb_2I_9$: A lead-free layered perovskite semiconductor. Chem Mater 2015;27:5622-5632.

[13] Fujihara T, Terakawa S, Matsushima T, Qin C, Yahiro M, Adachi C. Fabrication of high coverage $MASnI_3$ perovskite films for stable, planar heterojunction solar cells. J Mater Chem C 2017;5:1121-1127.

[14] Lyu M, Yun J-H, Cai M, Jiao Y, Bernhardt PV, Zhang M, Wang Q, Du A, Wang H, Liu G. Organic–inorganic bismuth (III)-based material: A lead-free, air-stable and solution-processable light-absorber beyond organolead perovskites. Nano Res 2016;9:692-702.

[15] Damjanovic D, Klein N, Li J, Porokhonskyy V. What can be expected from lead-free piezoelectric materials? Funct Mater Lett 2010;3:5-13.

[16] Nedelcu G, Protesescu L, Yakunin S, Bodnarchuk MI, Grotevent MJ, Kovalenko MV. Fast anion-exchange in highly luminescent nanocrystals of cesium lead halide perovskites ($CsPbX_3$, $X$= Cl, Br, I). Nano Lett 2015;15:5635-5640.

[17] Protesescu L, Yakunin S, Bodnarchuk MI, Krieg F, Caputo R, Hendon CH, Yang RX, Walsh A, Kovalenko MV. Nanocrystals of cesium lead halide perovskites ($CsPbX_3$, $X$= Cl, Br, and I): novel optoelectronic materials showing bright emission with wide color gamut. Nano Lett 2015;15:3692-3696.

[18] Slavney AH, Hu T, Lindenberg AM, Karunadasa HI. A bismuth-halide double perovskite with long carrier recombination lifetime for photovoltaic applications. J Am Chem Soc 2016;138:2138-2141.

[19] McClure ET, Ball MR, Windl W, Woodward PM. $Cs_2AgBiX_6$ ($X$= Br, Cl): new visible light absorbing, lead-free halide perovskite semiconductors. Chem Mater 2016;28:1348-1354.

[20] Luo J, Wang X, Li S, Liu J, Guo Y, Niu G, Yao L, Fu Y, Gao L, Dong Q. Efficient and stable emission of warm-white light from lead-free halide double perovskites. Nature 2018;563:541-545.

[21] Steele JA, Pan W, Martin C, Keshavarz M, Debroye E, Yuan H, Banerjee S, Fron E, Jonckheere D, Kim CW. Photophysical Pathways in Highly Sensitive $Cs_2AgBiBr_6$ Double-Perovskite Single-Crystal x-Ray Detectors. Adv Mater 2018;30:1804450.

[22] Xiao Z, Meng W, Wang J, Yan Y. Thermodynamic Stability and Defect Chemistry of Bismuth-Based Lead-Free Double Perovskites. ChemSusChem 2016;9:2628-2633.

[23] Hull S, Berastegui P. Crystal structures and ionic conductivities of ternary derivatives of the silver and copper monohalides-II: ordered phases within the $(AgX)_x$-$(MX)_{1-x}$ and $(CuX)_x$-$(MX)_{1-x}$ (*M*= K, Rb and Cs; *X*= Cl, Br and I) systems. J Solid State Chem 2004;177:3156-3173.

[24] Lazarini F. Caesium enneabromodibismuthate (III). Acta Crystallogr B 1977;33:2961-2964.

[25] Tubandt C. Über Elektrizitätsleitung in festen kristallisierten Verbindungen. Zweite Mitteilung. Überführung und Wanderung der Ionen in einheitlichen festen Elektrolyten. Z Anorg Allg Chem 1921;115:105-126.

[26] Tuller H. Ionic Conduction and Applications. In: Kasap S, Capper P, editors. Springer Handbook of Electronic and Photonic Materials. Cham: Springer; 2017, p. 247-263.

[27] Malugani J, Wasniewski A, Doreau M, Robert G, Al Rikabi A. Conductivite ionique dans les verres $AgPO_3$-Ag*X* (*X*= I, Br, Cl). Mater Res Bull 1978;13:427-433.

[28] Zurbuchen M, Lettieri J, Fulk S, Jia Y, Carim A, Schlom D, Streiffer S. Bismuth volatility effects on the perfection of $SrBi_2Nb_2O_9$ and $SrBi_2Ta_2O_9$ films. Appl Phys Lett 2003;82:4711-4713.

[29] Kim S, Choi E, Bhalla A. Effects of excess bismuth content in precursor solutions on ferroelectric properties of $BiFeO_3$ thin films prepared by a chemical solution deposition. Ferroelectrics Lett Sec 2007;34:84-94.

[30] Watanabe H, Mihara T, Yoshimori H, de Araujo CAP. Preparation of ferroelectric thin films of bismuth layer structured compounds. Jap J Appl Phys 1995;34:5240-5244.

[31] Pan W, Wu H, Luo J, Deng Z, Ge C, Chen C, Jiang X, Yin W-J, Niu G, Zhu L. $Cs_2AgBiBr_6$ single-crystal x-ray detectors with a low detection limit. Nat Photonics 2017;11:726.

[32] Zhang M, Zheng Z, Fu Q, Chen Z, He J, Zhang S, Yan L, Hu Y, Luo W. Growth and characterization of all-inorganic lead halide perovskite semiconductor $CsPbBr_3$ single crystals. CrystEngComm 2017;19:6797-6803.

[33] Furukawa Y, Sato M, Kitamura K, Yajima Y, Minakata M. Optical damage resistance and crystal quality of LiNbO$_3$ single crystals with various [Li]/[Nb] ratios. J Appl Phys 1992;72:3250-3254.

[34] Yamashita S, Kikkawa J, Yanagisawa K, Nagai T, Ishizuka K, Kimoto K. Atomic number dependence of Z contrast in scanning transmission electron microscopy. Sci Rep 2018;8:12325.

[35] Yang J, Zhang P, Wei S-H. Band structure engineering of Cs$_2$AgBiBr$_6$ perovskite through order–disordered transition: a first-principle study. J Phys Chem Lett 2017;9:31-35.

[36] Dong Q, Fang Y, Shao Y, Mulligan P, Qiu J, Cao L, Huang J. Electron-hole diffusion lengths> 175 μm in solution-grown CH$_3$NH$_3$PbI$_3$ single crystals. Science 2015;347:967-970.

[37] Chiu F-C. A review on conduction mechanisms in dielectric films. Adv Mater Sci Eng 2014;2014:578168.

[38] Jain A, Kumar P, Jain S, Kumar V, Kaur R, Mehra R. Trap filled limit voltage ($V_{TFL}$) and $V^2$ law in space charge limited currents. J Appl Phys 2007;102:094505.

[39] Lampert MA. Simplified theory of space-charge-limited currents in an insulator with traps. Phys Rev 1956;103:1648-1656.

[40] Mott NF, Gurney RW. Electronic processes in ionic crystals. J Chem Educ 1940;18:249.

[41] Yuan W, Niu G, Xian Y, Wu H, Wang H, Yin H, Liu P, Li W, Fan J. In Situ Regulating the Order–Disorder Phase Transition in Cs$_2$AgBiBr$_6$ Single Crystal toward the Application in an x-Ray Detector. Adv Funct Mater 2019;29:1900234.

[42] Sze SM, Ng KK. Physics of Semiconductor Devices. New Jersy: John Wiley & Sons; 2007.

[43] Lee JW, Kim SG, Yang JM, Yang Y, Park NG. Verification and mitigation of ion migration in perovskite solar cells. APL Mater 2019; 7: 041111.

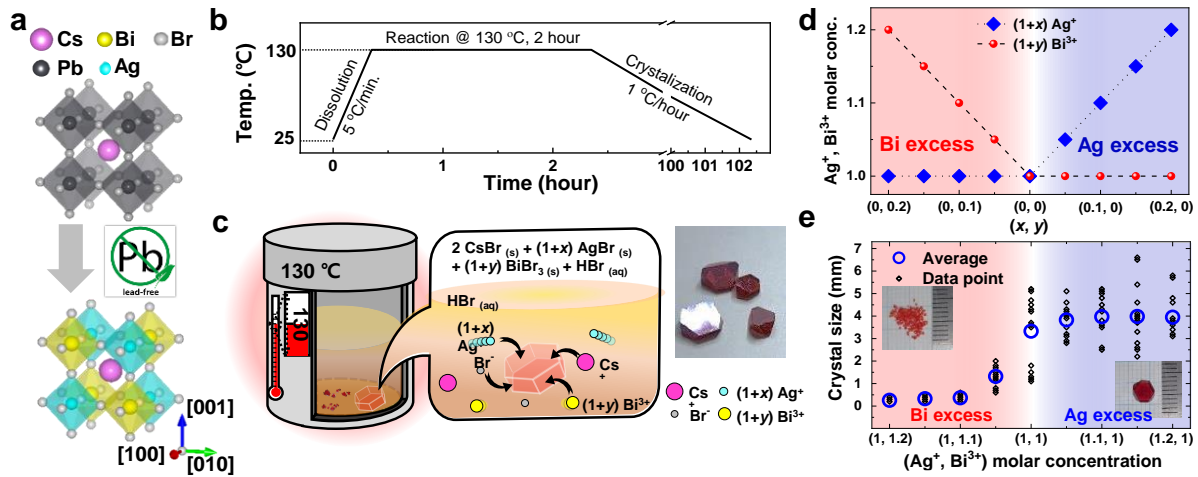

**Figure 1.** Lead-free double perovskite halide crystal growth. (a) Schematic view of lead-based perovskite halide CsPbBr$_3$ and lead-free double perovskite halide Cs$_2$AgBiBr$_6$. (b) Temperature-time sequence diagram for the hydrothermal synthesis of Cs$_2$AgBiBr$_6$. (c) A schematic diagram of a crystallization process for Cs$_2$AgBiBr$_6$ double perovskite formation and optical image of the crystals. (d) Initial molar concentrations of Ag$^+$ and Bi$^{3+}$ ions in the precursor solution to generate excess Ag or Bi conditions in a hydrothermal reaction. (e) The lateral sizes of Cs$_2$AgBiBr$_6$ crystals according to the initial molar concentrations of Ag$^+$ and Bi$^{3+}$ ions in a hydrothermal reaction.

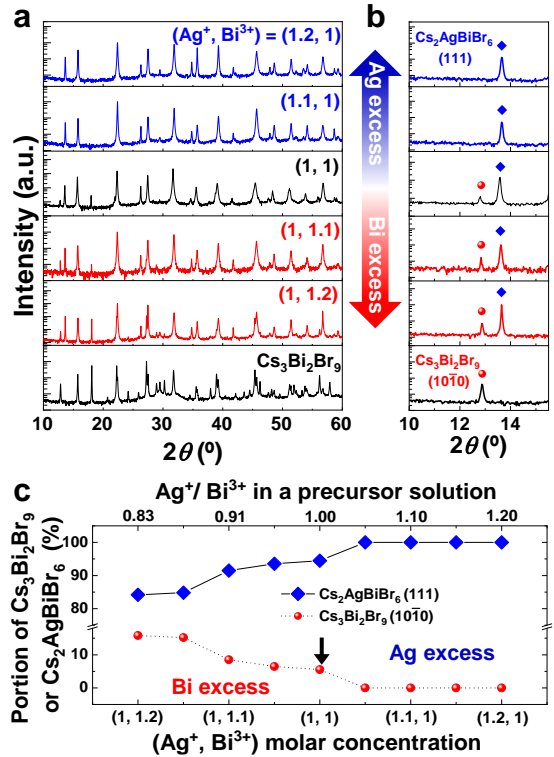

**Figure 2.** Structure characterization. (a) X-ray diffraction patterns of ground powder of $Cs_2AgBiBr_6$ single crystals grown under various chemical environments (i.e., stoichiometric, excess Bi, and excess Ag conditions), and $Cs_3Bi_2Br_9$ powder (as a reference for the secondary phase). (b) Enlarged view of the XRD peak corresponding to the (111) Bragg peak of $Cs_2AgBiBr_6$ and ($10\bar{1}0$) Bragg peak of $Cs_3Bi_2Br_9$ around a $2\theta$ angle of 13°. (c) The volume fractions between the primary $Cs_2AgBiBr_6$ and secondary $Cs_3Bi_2Br_9$ phases based on the initial molar concentrations of $Ag^+$ and $Bi^{3+}$ ions in a hydrothermal reaction.

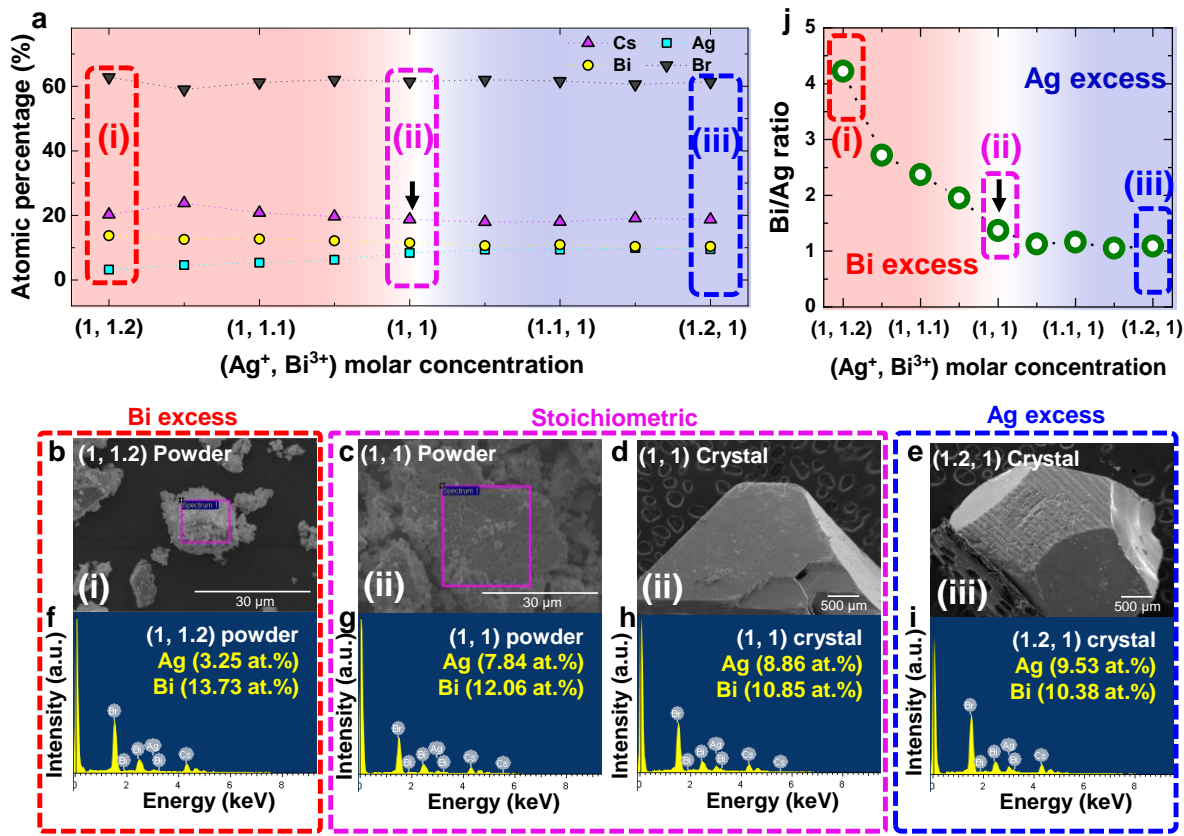

**Figure 3** Composition analysis. (a) Atomic composition as a function of the initial molar concentrations of $Ag^+$ and $Bi^{3+}$ ions in the as-grown $Cs_2AgBiBr_6$ powders and single crystals via a hydrothermal reaction. (b, c, d, e) SEM images and (f, g, h, i) EDX spectra of $Cs_2AgBiBr_6$ powders and crystals grown under various chemical environments [i.e., Bi-excess (b, f), stoichiometric (c, d, g, h), and Ag-excess (e, i) conditions]. (j) The estimated molar ratio of Bi/Ag as a function of the initial molar concentrations of $Ag^+$ and $Bi^{3+}$ ions in as-grown $Cs_2AgBiBr_6$ samples.

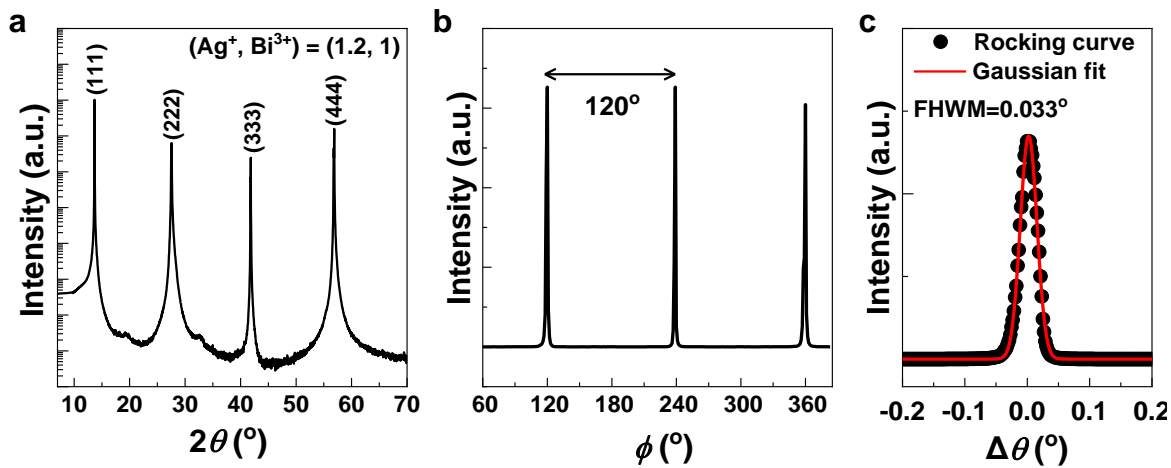

**Figure 4.** Crystallinity of our $Cs_2AgBiBr_6$ single crystal grown under an Ag-excessive environment. The high-resolution XRD data of (a) the $\theta$-$2\theta$ scan, (b) the phi ($\phi$) scan, and (c) the rocking-curve measurement of the $Cs_2AgBiBr_6$ single crystal grown under an Ag-excessive environment.

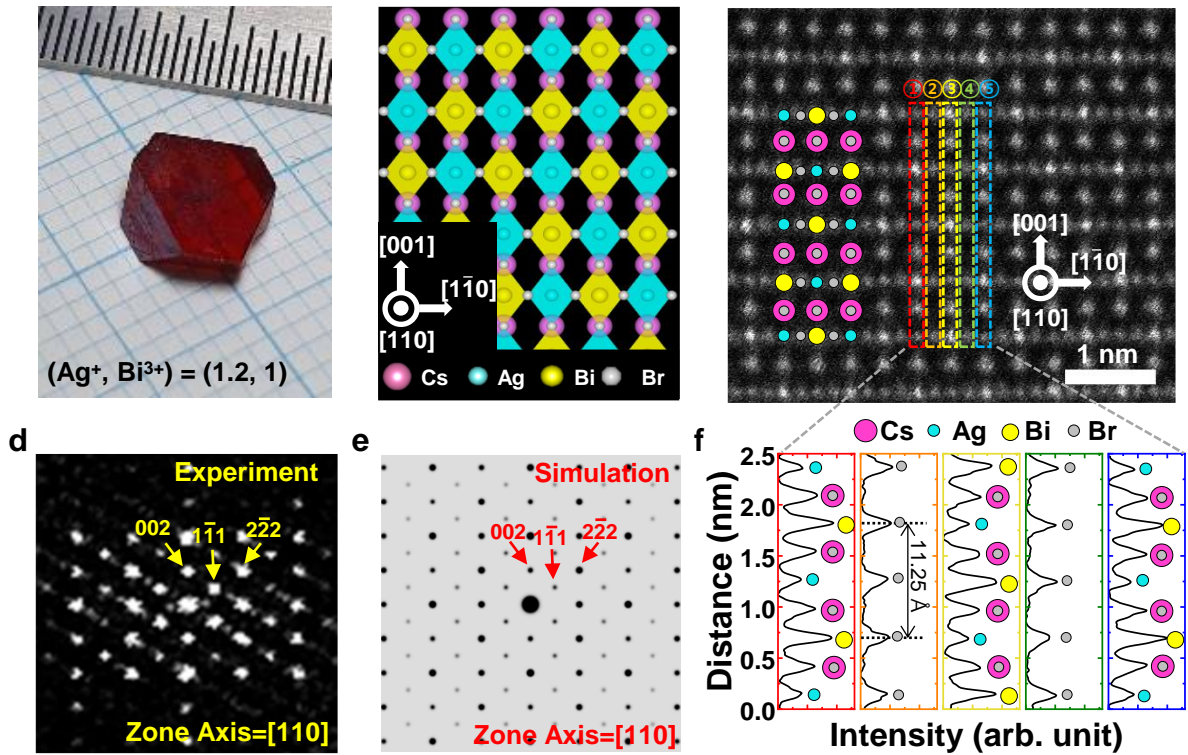

**Figure 5.** Atomic-resolution STEM images of a $Cs_2AgBiBr_6$ single crystal grown under an Ag-excess environment. (a) Optical image of $Cs_2AgBiBr_6$ single crystals. (b) Schematic diagram of the $Cs_2AgBiBr_6$ structure viewed along the [110] direction. (c) High-resolution HAADF-STEM image, (d) FFT pattern of the $Cs_2AgBiBr_6$ single crystals, and (e) simulated electron diffraction pattern of a cubic double perovskite structure along the [110] zone axis. (f) HAADF-STEM intensity profiles along the [001] direction from regions shown in (c) corresponding to the atomic rows shown in the schematic.

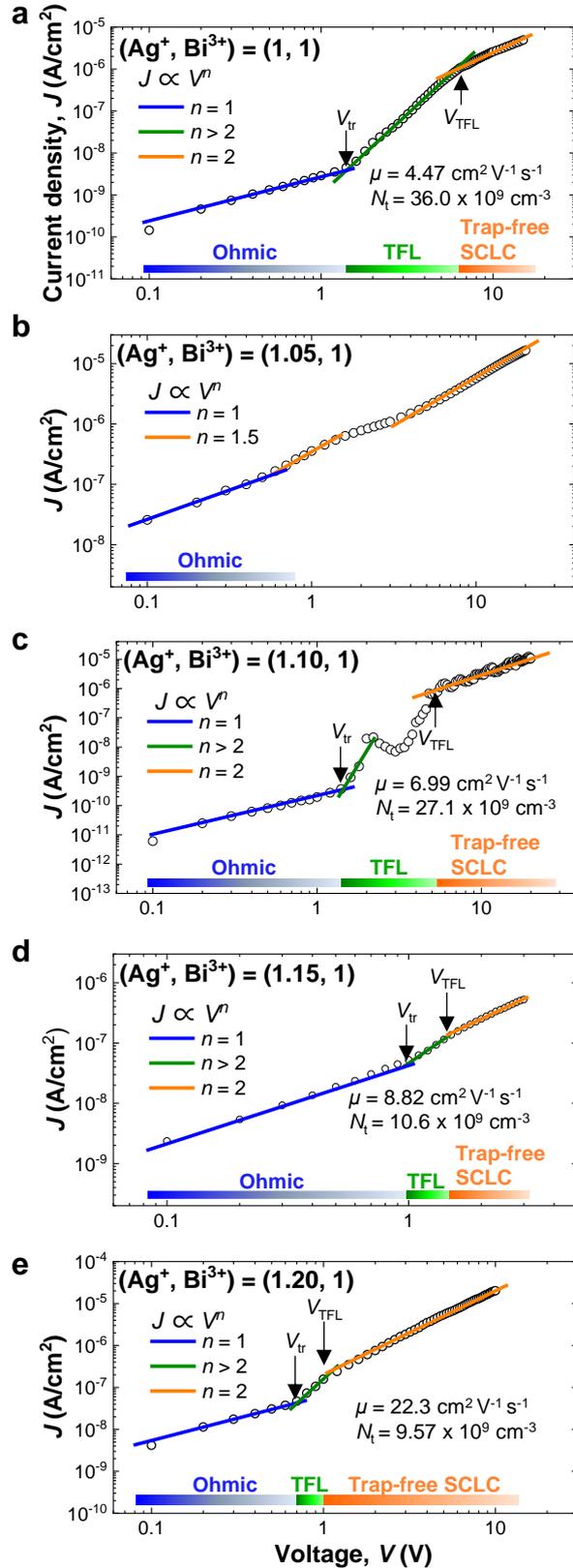

**Figure 6.** Carrier mobility characterization. Logarithm J–V curves in the dark for $Cs_2AgBiBr_6$ single crystals grown under (a) stoichiometric and (b, c, d, e) Ag-excessive conditions. Linear fittings are applied to estimate the carrier mobility and trap density according to the space charge-limited current (SCLC) model. The regions are marked for Ohmic (Blue, $J \propto V^{n=1}$), TFL (trap-filled limited) (Green, $J \propto V^{n>2}$) and Trap-free SCLC regime (Orange, $J \propto V^{n=2}$).

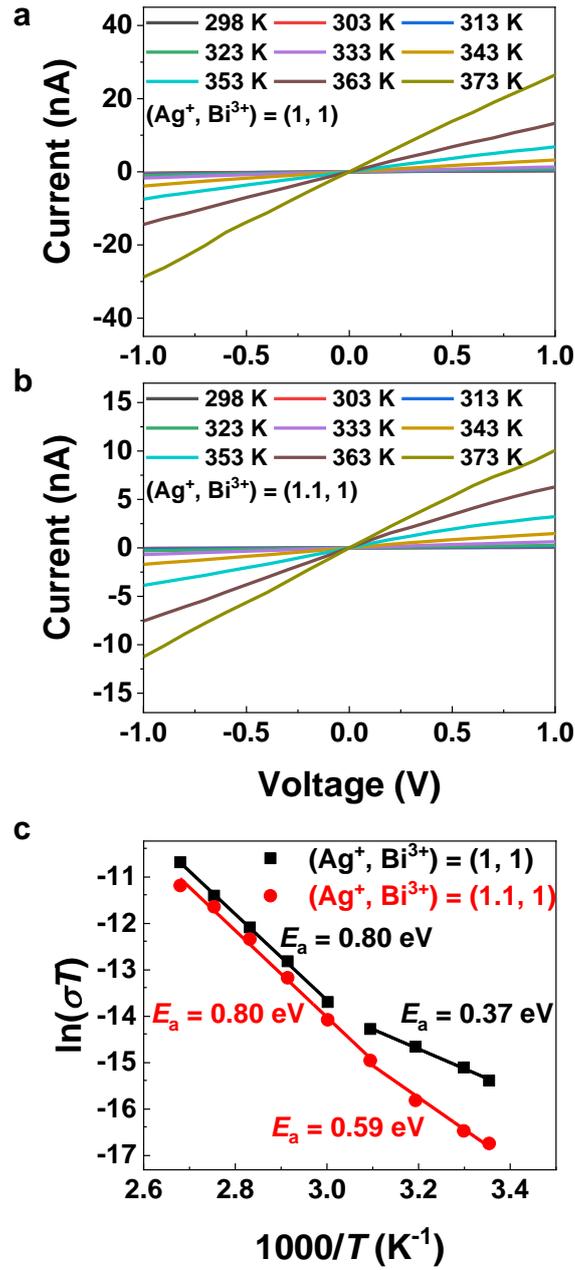

**Figure 7.** (a), (b) Temperature-dependent dark current of Cs$_2$AgBiBr$_6$ crystal grown under various chemical environments (i.e., stoichiometric (a) and Ag-excess conditions (b)). (c) Arrhenius plots of the temperature dependence of $\sigma T$ versus $1000/T$.

**Table 1.** Formation energies of Ag and Bi vacancy defects from first-principles calculations.

| Compound | Ag vacancy | Bi vacancy |
|---|---|---|
| $E_{vf}$ (eV) | 3.13 | 6.15 |

**Supporting Information**

**Highly ordered lead-free double perovskite halides by design**

Chang Won Ahn[a,†], Jae Hun Jo[a,†], Jong Chan Kim[b], Hamid Ullah[c], Sangkyun Ryu[d], Younghun Hwang[e], Jin San Choi[a], Jongmin Lee[f], Sanghan Lee[f], Hyoungjeen Jeen[d], Young-Han Shin[c], Hu Young Jeong[b], Ill Won Kim[a], and Tae Heon Kim[a]*

[a]*Department of Physics and Energy Harvest-Storage Research Center (EHSRC), University of Ulsan, Ulsan 44610, Republic of Korea.*

[b]*UNIST Central Research Facilities (UCRF) & School of Materials Science and Engineering, Ulsan National Institute of Science and Technology (UNIST), Ulsan 44919, Republic of Korea.*

[c]*Multiscale Materials Modelling Laboratory, Department of Physics, University of Ulsan, Ulsan 44610, Republic of Korea.*

[d]*Department of Physics, Pusan National University, Pusan 46241, Republic of Korea.*

[e]*School of Electrical and Electronics Engineering, Ulsan College, Ulsan 44610, Republic of Korea.*

[f]*School of Materials Science and Engineering, Gwangju Institute of Science and Technology, Gwangju 61005, Republic of Korea.*

*Electronic mail: thkim79@ulsan.ac.kr

[†]These authors contributed equally to this work.

Keywords: $Cs_2AgBiBr_6$, lead-free, double perovskite, single crystal.

**Starting materials.** Cesium bromide [CsBr] (99.9%), silver bromide [AgBr] (99%), bismuth bromide [BiBr$_3$] (≥ 98%), lead bromide [PbBr$_2$] (≥ 98%), hydro bromic acid [HBr] (48% in H$_2$O), N,N-dimethyl form amide [DMF] (99.8%), diethyl ether (≥ 99%), and ethyl alcohol anhydrous (99.9%) were purchased from Sigma Aldrich Co. LLC. Methylamine (CH$_3$NH$_2$) (40 wt. % in methanol) was purchased from Tokyo Chemical Industry Co., Ltd.

**Crystallization of Cs$_2$AgBiBr$_6$.** To fabricate Cs$_2$AgBiBr$_6$ single crystals, we used a conventional hydrothermal reaction technique. For the hydrothermal reaction, we first prepared a starting solution by dissolving CsBr, AgBr, and BiBr$_3$ powders in the HBr solvent. The solution concentration was controlled at 0.1 M. In the case of the stoichiometric molar ratio, 0.002 mol CsBr (0.426 g), 0.001 mol AgBr (0.188g), and 0.001 mol BiBr$_3$ (0.449 g) were dissolved into 10 mL HBr solution. And, the solution was put into a 50 mL Teflon-lined stainless-steel autoclave and fired in a box furnace up to 130 ˚C with the ramping rate of 5 ˚C/min. Then, it was cooled down to room temperature with the ramping rate of 1 ˚C/hour for crystallization. Finally, the products were filtered through a glass fiber membrane filter, washed with diethyl ether and dried in a vacuum furnace at 60 ºC for 1 day. To control the initial chemical environment (Ag-excess or Bi-excess) for reaction in the hydrothermal synthesis of Cs$_2$AgBiBr$_6$ single crystals, the starting solutions for a hydrothermal reaction were prepared according to the Equation 2.

**Crystallization of Cs$_3$Bi$_2$I$_9$.** Cs$_3$Bi$_2$I$_9$ crystals were grown using hydrothermal synthesis to identify the secondary phase produced when preparing Cs$_2$AgBiBr$_6$ single crystals. CsBr and BiBr$_3$ (3/2 by molar, 0.1 M) were dissolved in the HBr solvent. And, the solution was put into a 50 mL Teflon-lined stainless-steel autoclave and fired in a box furnace up to 130 ˚C with the ramping rate of 5 ˚C/min. Then, it was cooled down to room temperature with the ramping rate of 1 ˚C/hour for crystallization. Finally, the products were filtered through a glass fiber membrane filter, washed with diethyl ether and dried in a vacuum furnace at 60 ºC for 1 day.

**CH$_3$NH$_3$Br precursors synthesis.** The methyl ammonium bromide (CH$_3$NH$_3$Br) precursors were synthesized from hydro bromide acid HBr with methylamine in round-bottomed flask at 0 °C for 2 h with stirring. The resulting solutions were recovered by evaporating at 80 °C for 24 h. The products were dissolved in ethyl alcohol anhydrous at room temperature, and then recrystallized using diethyl ether, and the precipitate filtered using vacuum filter. Finally, the yield white $CH_3NH_3Br$ powder dried at 60 °C in vacuum oven for 24 h.

**Crystallization of $CH_3NH_3PbBr_3$.** To grow the single crystalline perovskite $CH_3NH_3PbBr_3$, the solution concentration was controlled at 1.23 M in DMF with an equimolar mixture of the $CH_3NH_3Br$ and $PbBr_2$. The mixtures were dissolved at 50 °C for 2 h with stirring and then, filtered by PTFE membrane filter (0.2 μm pore size). The contained solution in petri dish was kept in oil bath on hot plate at 100 °C during several days. After a chemical reaction, the large number of small (~1-2 mm in size) seed crystals was fabricated and they were visible. For the further growth of single crystals with larger size, two or three seed crystals were inserted into the precursor solution repeatedly. A few hours later, the seed crystals were crystallized to $CH_3NH_3Br$ single crystals with the lateral size of 5-6 mm.

**Powder and single crystal X-ray diffraction.** Synchrotron x-ray diffraction (XRD) experiments were performed with the 3A beam line of Pohang Accelerator Laboratory (PAL). In the XRD measurements, a 6-circle x-ray goniometer was used for the XRD $\theta$-$2\theta$ scans, the phi ($\phi$) scans, and rocking-curve measurements.

**FE-SEM measurements and EDX elemental analysis.** A field emission scanning electron microscope (FE-SEM, JSM-7600, JEOL, Japan) with an energy-dispersive X-ray spectroscopy (EDX) detector was used to visualize the morphology and analyze the chemical composition of the as-synthesized $Cs_2AgBiBr_6$ single-crystal/powder compounds.

**Scanning transmission electron microscope (STEM) measurement.** The cross-sectional TEM samples of $Cs_2AgBiBr_6$ single crystals were prepared with the use of focused ion beam (FEI Helios Nano Lab 450) milling technique. To reduce air-exposure time of lamella samples, they were directly transferred to TEM column from FIB chamber in a minute. Atomic-resolution HAADF-STEM imaging and EDX elemental mapping were performed using an aberration-corrected STEM (FEI Titan[3] G2 60-300) running at an operation voltage of 200 kV, equipped with a Super-X EDX detector system. The probe convergence semi-angle was set to be approximately 25 mrad. HAADF-STEM images were acquired over a detector angle range of 50-200 mrad. To reduce the electron-beam damage, we used a low probe current of < 30 pA.

**Optical properties.** The sample for optical absorption measurements was approximately 500 μm thick. Optical absorption was measured using a tungsten halogen lamp (500 W) and a spectrometer (DP320i, Dongwoo optron Co. Ltd.) at room temperature. The absorption edge ($E_g$) of the sample was evaluated from the absorption coefficient using the Tauc equation $(\alpha h\nu) \sim (h\nu - E_g)^n$.[1] Note that the exponent $n$ is characterized by the type of optical transition, where $n$ are 1/2 and 2 for direct indirect transitions, respectively. In order to determine $E_g$, $(\alpha h\nu)^{1/n}$ is plotted as a function of $h\nu$. Here, it is possible to derive the $E_g$ value by extrapolating the curve as a linear plot and thereafter, extracting the photon energy intercept.

**J-V measurements (SCLC).** J-V curve was measured using a Keithley 237, using a metal-insulator-metal (MIM) structure of Au/Cs$_2$AgBiBr$_6$/Au for the hole-governing electrical transport measurements. The sample was kept in a dark environment at monitored room temperature. The applied bias at the kink point between trap-filled limited region and trap-free space-charge-limited region is known as the trap-filled limit voltage ($V_{TFL}$), which allows us to determine the trap density $N_t = V_{TFL}(2\varepsilon\varepsilon_0)/(eL^2)$ from the Equation 4.[2-5] And, the carrier mobility is also computed to $\mu = 8L^3J/(9\varepsilon\varepsilon_0V^2)$ from the Equation 3.[2-5]

**Dielectric constant measurement.** To measure the dielectric constant, we prepared a simple parallel-plate capacitor by deposition of Au electrodes on both sides of Cs$_2$AgBiBr$_6$ single crystals with flat surfaces. We measured the frequency dependent capacitance of Cs$_2$AgBiBr$_6$ single-crystal capacitors at dark using HP4192A impedance analyzer. And, we obtained the relative dielectric constant ($\varepsilon$) of Cs$_2$AgBiBr$_6$ using the parallel plate capacitor model:

$C = \varepsilon\varepsilon_0 A/d$ (5),

where $C$ is capacitance of single crystals, $\varepsilon_0$ is vacuum permittivity, $A$ is the electrode area, $d$ is the thickness of single crystals.

**Characterization of the activation energy for electrical transport property in Cs$_2$AgBiBr$_6$ single crystal.** To measure the I-V curve we prepared a simple parallel-plate capacitor by deposition of Au electrodes on both sides of Cs$_2$AgBiBr$_6$ single crystals with flat surfaces. We measured the voltage dependent current of $Cs_2AgBiBr_6$ single crystal capacitors at dark on varying temperatures using Keithley 237. And, we calculated the conductivity ($\sigma$) of $Cs_2AgBiBr_6$ using the equation:

$$\sigma = Id/AV \qquad (6),$$

where $\sigma$ is conductivity at given absolute temperature $T$, $A$ is the electrode area, $d$ is the thickness of single crystals.

To quantitatively characterize the electrical transport property, we obtained the activation energy by fitting $\ln(\sigma T)$ vs. $1/T$ using a Equation 7 with temperature-dependent conductivity.[14] The relationship between the activation energy and conductivity can be described by a following Arrhenius type equation given by

$$\sigma = \frac{\sigma_0}{T}\exp\left(\frac{-E_a}{k_B T}\right) \qquad (7),$$

where $\sigma$ is the conductivity at given absolute temperature $T$, $k_B$ is the Boltzmann's constant, $E_a$ is the activation energy, and $\sigma_0$ is the pre-exponential factor.

**Theoretical calculations.** We have performed the first-principles calculations in the framework of density functional theory (DFT) as implemented in the Vienna *ab-initio* Simulation Package (VASP).[6,7] The generalized gradient approximation (GGA) within the Perdew-Burke-Ernzerhof (PBE)[8] formalism is employed for the exchange-correlation potential. A cutoff energy of 520 eV is used in the calculations. For geometry optimization the Brillouin-zone integration is performed using $3 \times 3 \times 3$ k mesh within the Monkhorst-scheme. The convergence criterion of the self-consistent field calculations is set to $10^{-6}$ eV for the total energy. By using the conjugate gradient method, atomic positions and lattice constants are optimized until the Hellmann-Feynman forces are less than 0.001 eV/Å. However, it is well known that GGA underestimate the band gap.[9] To overcome the underestimation in electronic band gap value, we have used the hybrid functional Heyd-Scuseria-Ernzerh of (HSE06) calculations.[10] We employ HSE06 functional with the default screening parameter 0.2 Å$^{-1}$ as implemented in VASP. A super cell approach has been adopted to introduce a vacancy in $Cs_2AgBiBr_6$. For instance, one Ag or Bi atom is removed from the host 2×2×1 super cell to make a vacancy.

**Electronic properties.** We have calculated the electronics properties of $Cs_2AgBiBr_6$ in terms of calculating its band structure and density of states (DOS). We have performed the band structure calculation to investigate the electronic structure of the studies compounds. In contrast to the direct band gap nature of lead halide perovskite, such as $CsPbCl_3$, the double perovskite, $Cs_2AgBiBr_6$ possesses an indirect band gap nature.[11] The indirect band gap for $Cs_2AgBiBr_6$ is estimated to be 2.26 eV, which is in agreement with the previous reports.[11-13] The fundamental band gap arises due to the transition from the top of the valence band located at X to the bottom of conduction band located at Γ symmetry point, as shown in Figure S10a. To further insight into the electronic properties, we have calculated the total and projected DOS. The DOS of $Cs_2AgBiBr_6$, as depicted in Figure S10b, matches well with the band structure (See Figure S10a). Figure S10b shows that the valence band maximum is dominated by Ag-$d$ states and Br-$p$ states. The conduction band minimum is mostly composed of Bi-/Br-$p$ states and a small contribution comes from Ag-$s$ states. The transition, valence to conduction band, is mainly from the filled Br-$p$ states to anti-bonding Ag-$s$ and Bi-$p$ states.

**Vacancy formation.** The vacancy can be formed in $Cs_2AgBiBr_6$ by removing one species of Ag or Bi from the host supercell. In order to find how much energy is required to create the vacancy at an Ag/Bi site in $Cs_2AgBiBr_6$, we have calculated the vacancy formation energy ($E_{vf}$) using the relation, $E_{vf} = E_{total} - E_0 + E_{Ag/Bi}$, where $E_{total}$ and $E_0$ are the energies of $Cs_2AgBiBr_6$ with and without vacancy at the Ag-/Bi-site, respectively, and $E_{Ag/Bi}$ is the energy of a single Ag/Bi atom. The $E_{vf}$ values are tabulated in Table 1. Note that the $E_{vf}$ of the Ag vacant site is smaller than that of the Bi vacant site, which indicates that Ag vacancies in $Cs_2AgBiBr_6$ can be easily formed compared with Bi vacancies.

.

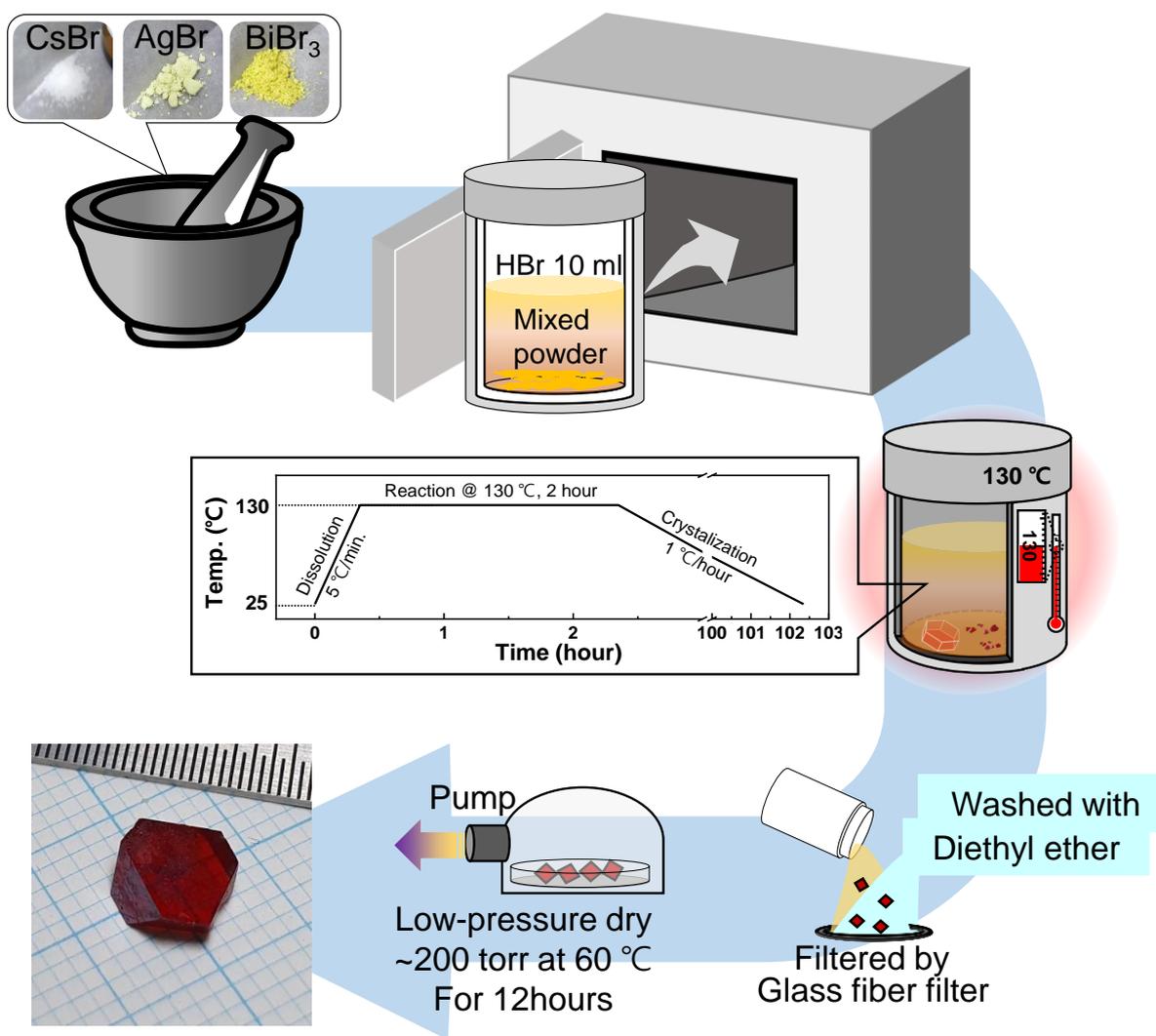

**Figure S1.** A schematic diagram of the Cs$_2$AgBiBr$_6$ single crystal growth by hydrothermal reaction process.

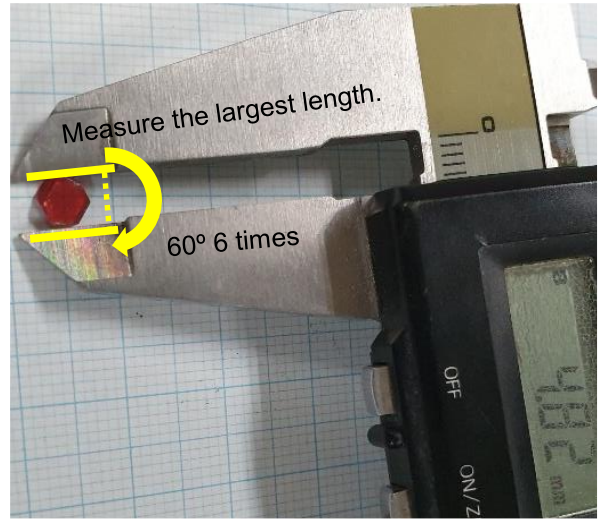

**Figure S2.** Size estimation of the as-grown $Cs_2AgBiBr_6$ single crystals using digital Vernier calipers.

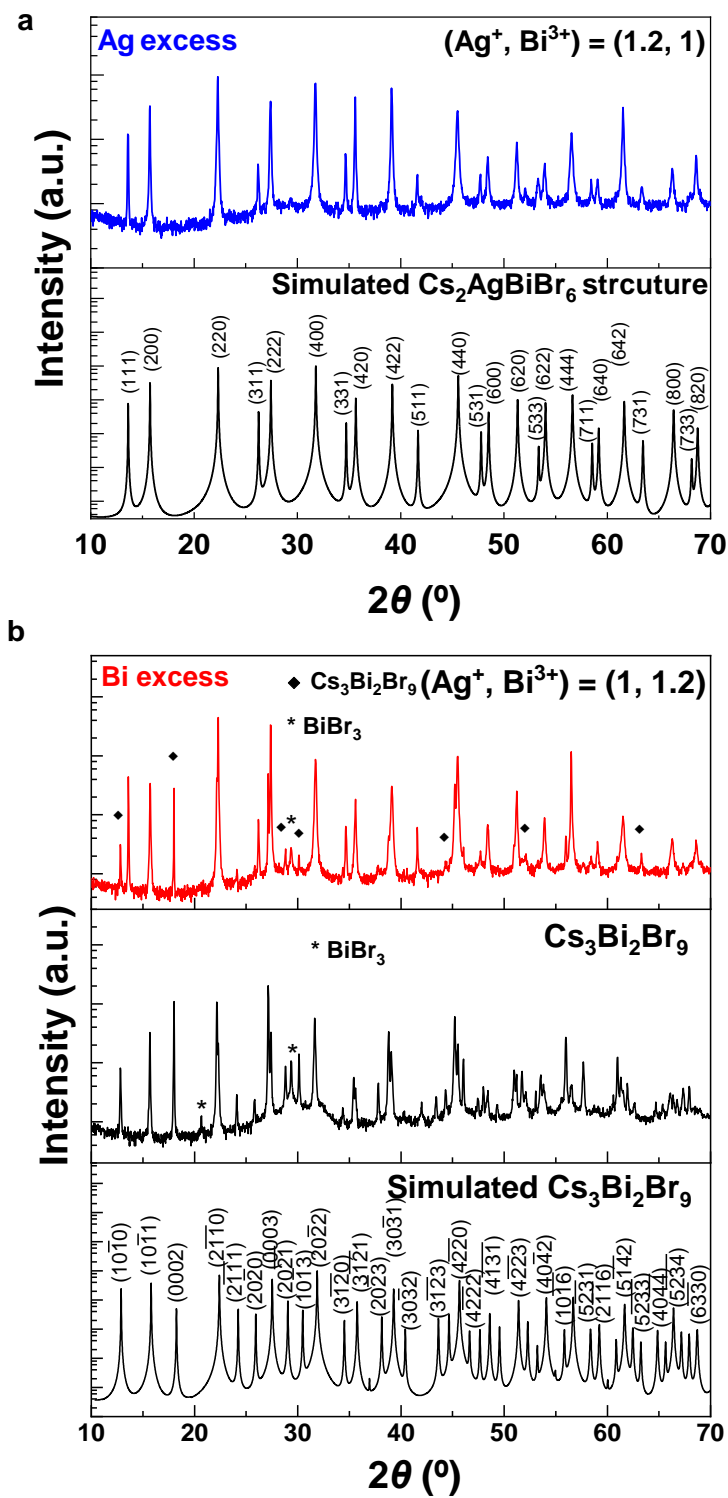

**Figure S3.** (a) XRD patterns of ground powder of $Cs_2AgBiBr_6$ single crystals grown under Ag-excess chemical environment and simulated XRD pattern of $Cs_2AgBiBr_6$ crystal structure (for the reference of primary phase). (b) XRD patterns of $Cs_2AgBiBr_6$ crystals grown under Bi-excess chemical environment, $Cs_3Bi_2Br_9$ powder, and simulated XRD pattern of $Cs_3Bi_2Br_9$ crystal structure (for the reference of secondary phase). Small XRD peaks of $BiBr_3$ residues are marked with an asterisk. The simulation was done by Visualization for Electronic and Structural Analysis (VESTA Ver. 3.4.4) program using raw data files ("mp-1078250" for $Cs_2AgBiBr_6$, "mp-27544" for $Cs_3Bi_2Br_9$) from "Materials Project" (materialsproject.org).

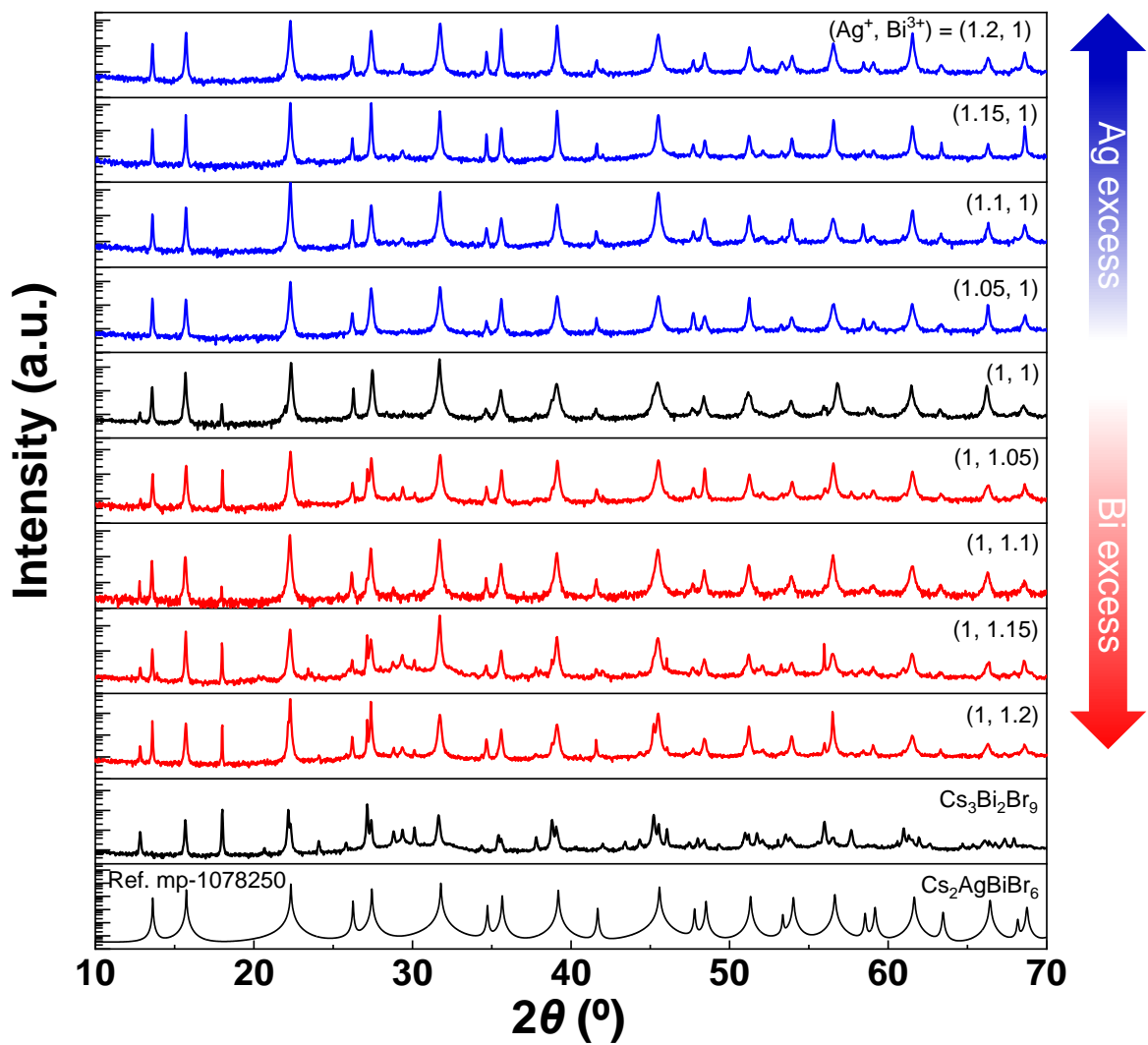

**Figure S4.** X-ray diffraction patterns of ground powder of $Cs_2AgBiBr_6$ single crystals, grown under various chemical environments (i.e., stoichiometric, Bi-excess, and Ag-excess conditions), $Cs_3Bi_2Br_9$ powder (for the reference of secondary phase), and simulated XRD pattern of $Cs_2AgBiBr_6$ structure (for the reference of primary phase).

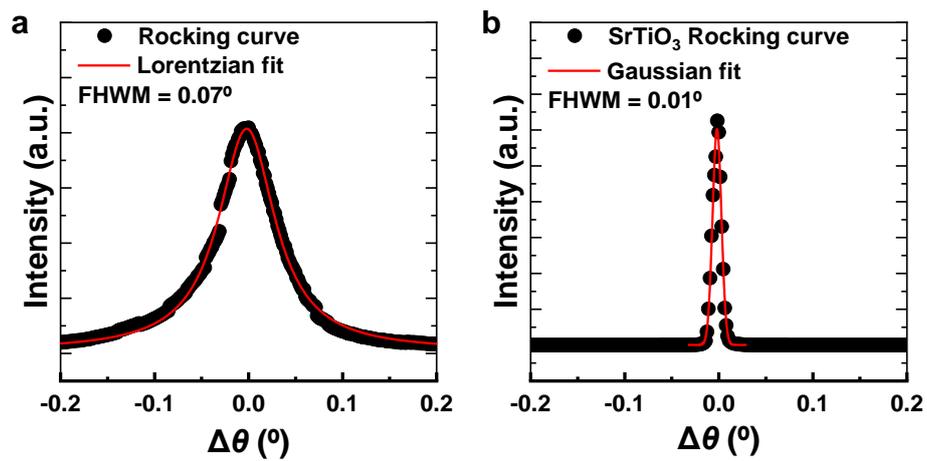

**Figure S5.** Rocking-curve measurement of (a) the MAPbBr$_3$ (MA$^+$=CH$_3$NH$_3{}^+$) and (b) SrTiO$_3$ single crystals.

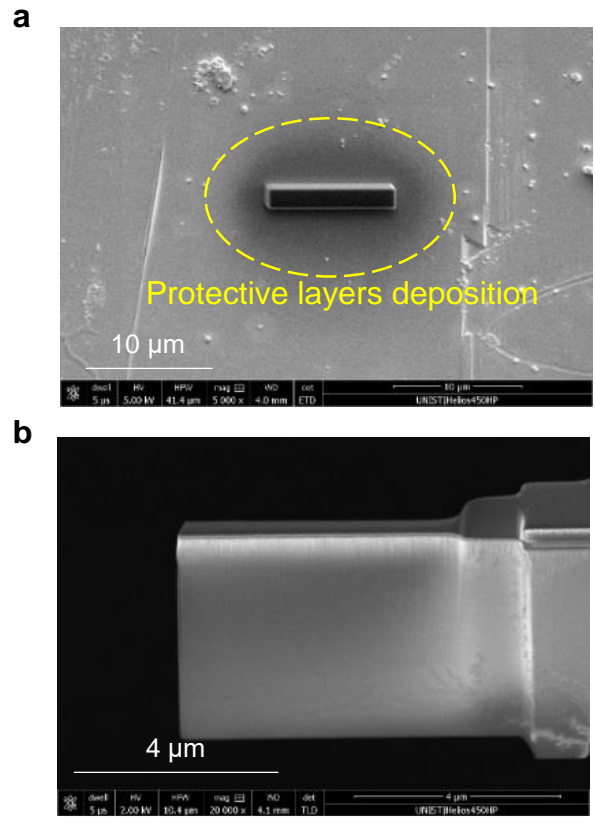

**Figure S6.** SEM image of cross-sectional TEM sample of $Cs_2AgBiBr_6$ single crystal grown under Ag-excess condition. (a) SEM images of protective layers (carbon/Pt) deposited surface and (b) cross-sectional TEM sample prepared by focused ion beam (FIB).

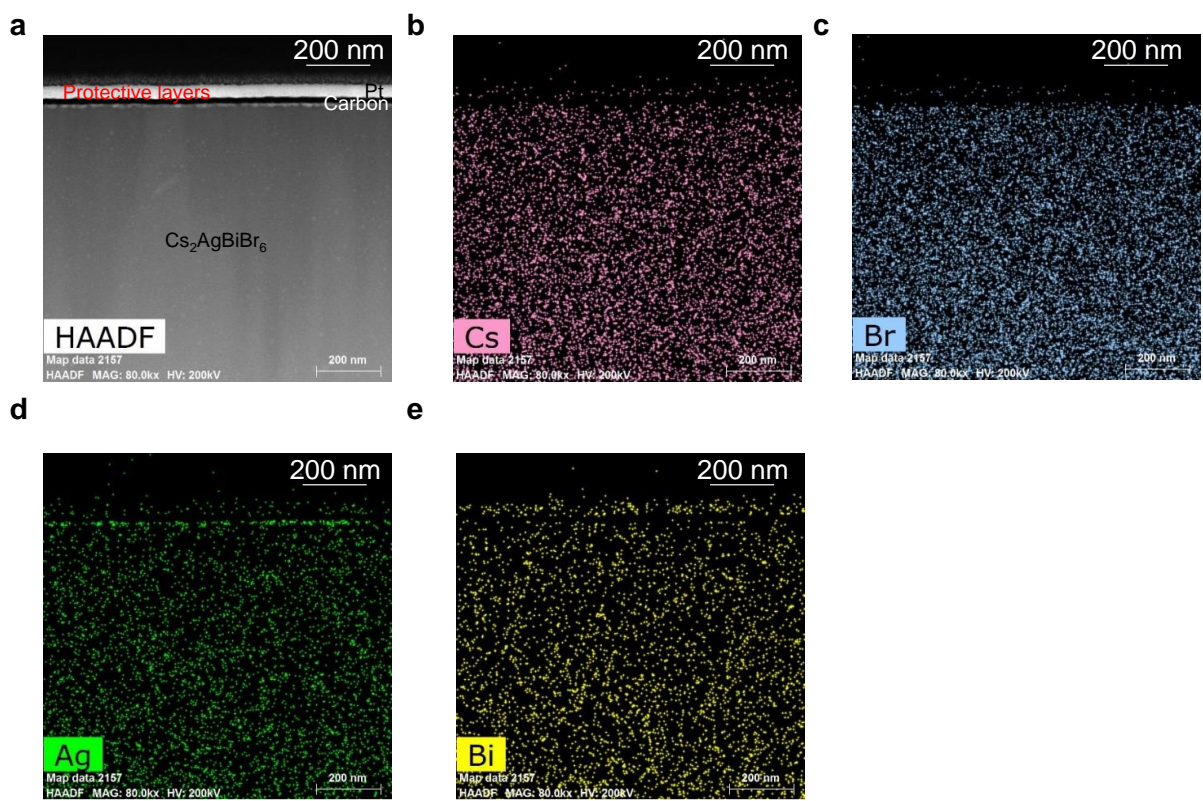

**Figure S7.** HAADF image and EDX elemental maps of cross-sectional TEM sample of Cs$_2$AgBiBr$_6$ crystal grown under Ag-excess condition. (a) HAADF image of Cs$_2$AgBiBr$_6$ crystal and corresponding EDX elemental maps for (b) Cs, (c) Br, (d) Ag, and (e) Bi.

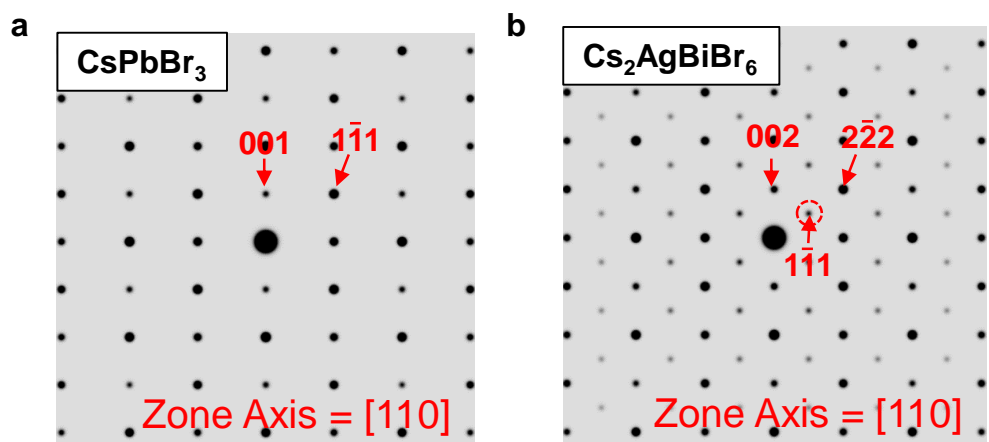

**Figure S8.** A comparison between the simulated electron diffraction patterns of (a) cubic perovskite (CsPbBr₃) and (b) double perovskite (Cs₂AgBiBr₆) structures.

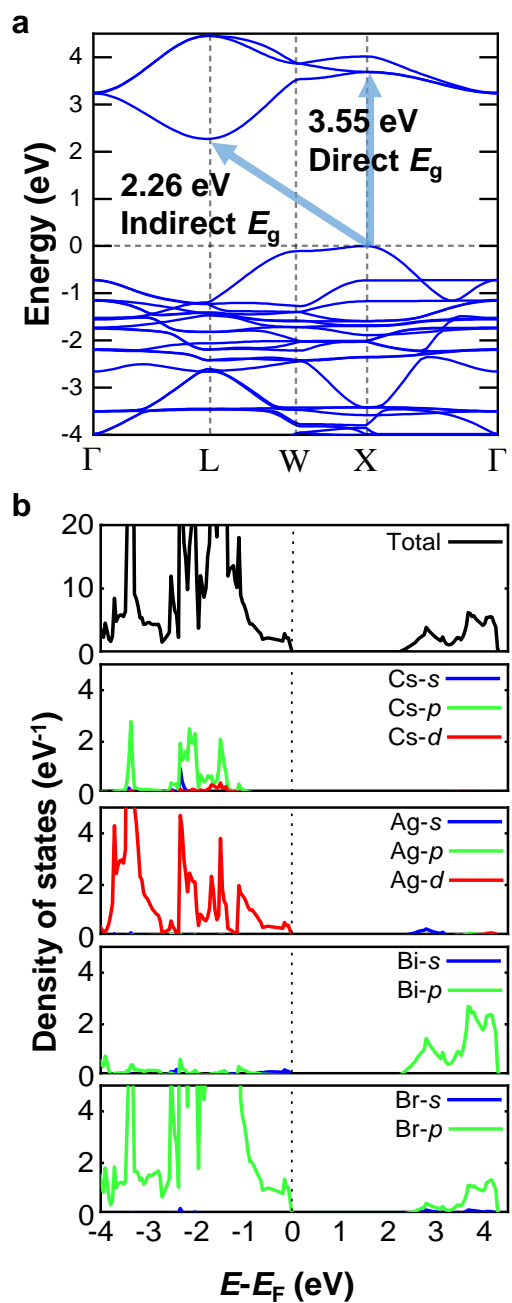

**Figure S9.** (a) The calculated band structures and (b) total and projected DOS of $Cs_2AgBiBr_6$ using HSE06 potential. The Fermi energy is set to zero and denoted by the horizontal dashed (band structure) and dotted vertical (DOS) line.

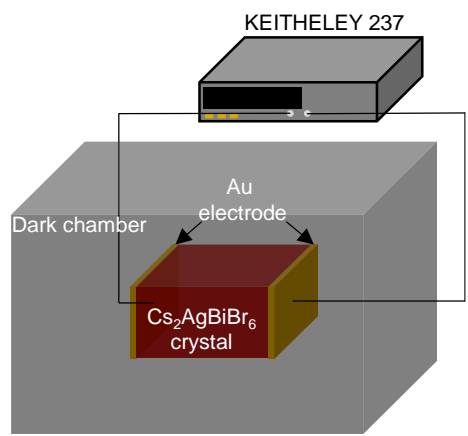

**Figure S10.** A schematic diagram of current density-voltage (*J-V*) measurements in the dark for the hole-governing electrical transport using Au/Cs$_2$AgBiBr$_6$/Au device.

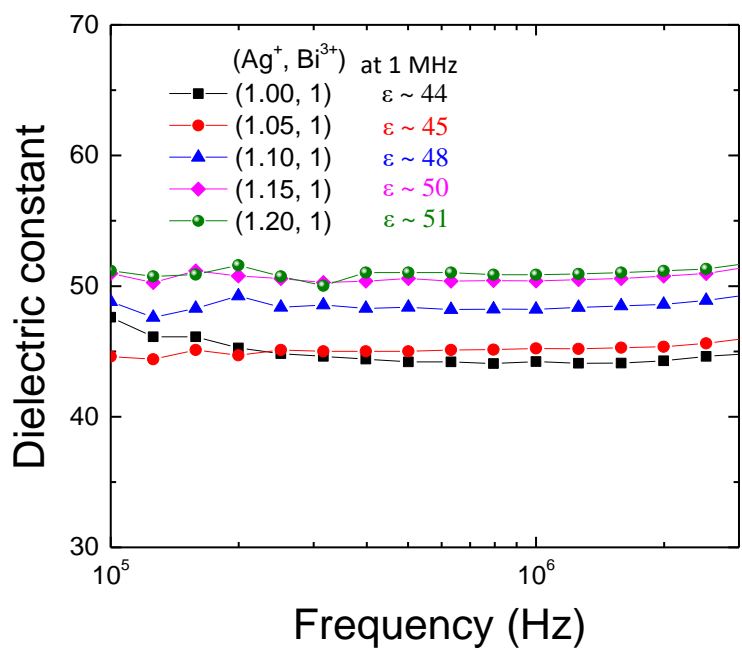

**Figure S11.** Frequency dependent dielectric constant for Cs$_2$AgBiBr$_6$ single crystals according to the initial molar concentrations of Ag$^+$ and Bi$^{3+}$ ions in a hydrothermal reaction.

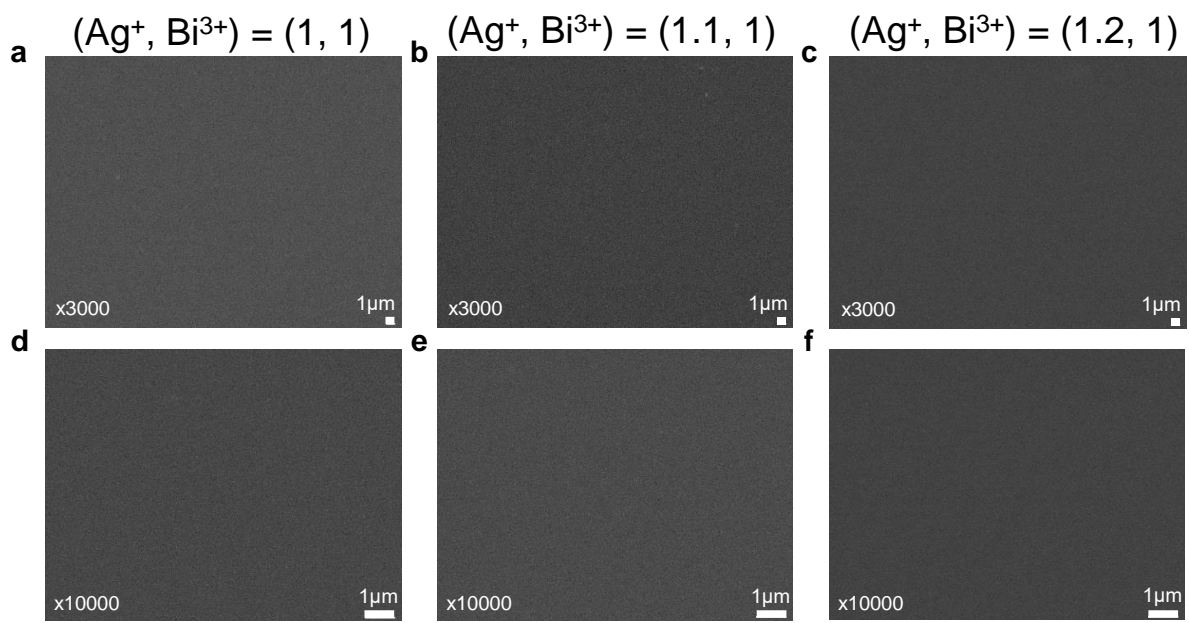

**Figure S12.** (a, b, c, d, e, f) SEM images of $Cs_2AgBiBr_6$ crystal surface grown under various chemical environments [i.e., stoichiometric (a, d) and Ag-excess conditions (b, c, e, f)].

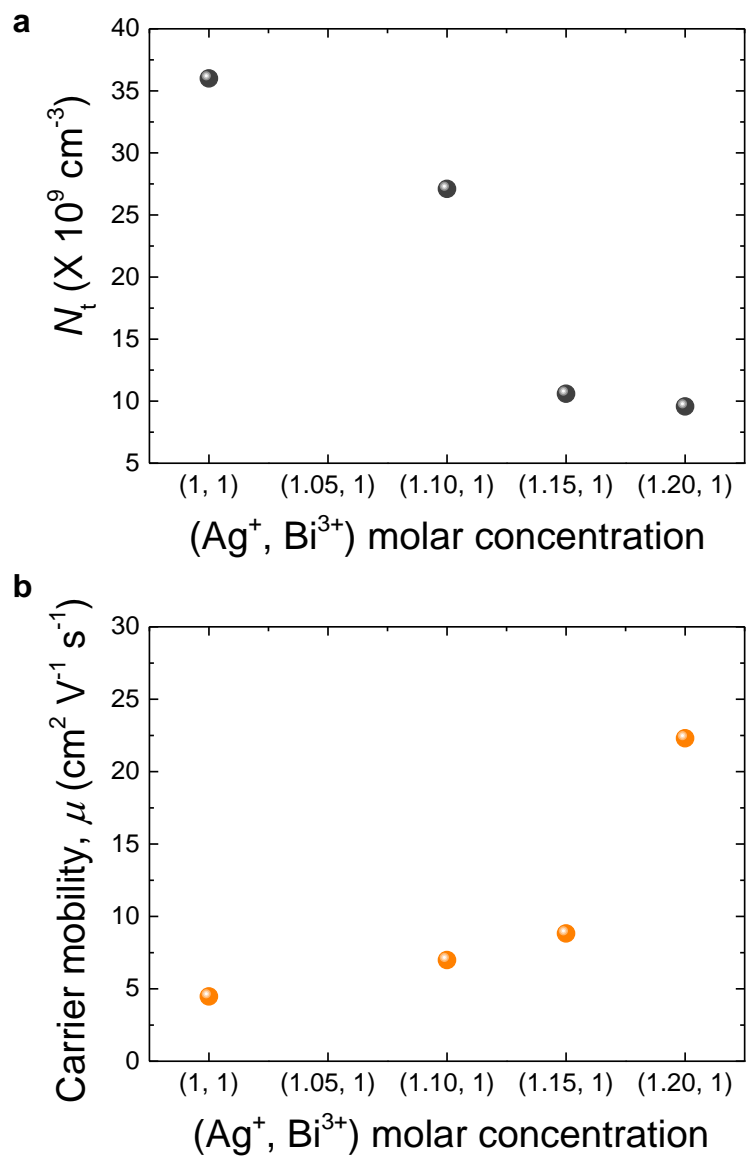

**Figure S13.** (a) Trap density ($N_t$) and (b) carrier mobility ($\mu$) according to the initial molar concentrations of $Ag^+$ and $Bi^{3+}$ ions in a hydrothermal reaction by fitting the measured *J-V* curves as shown in Fig. 6.

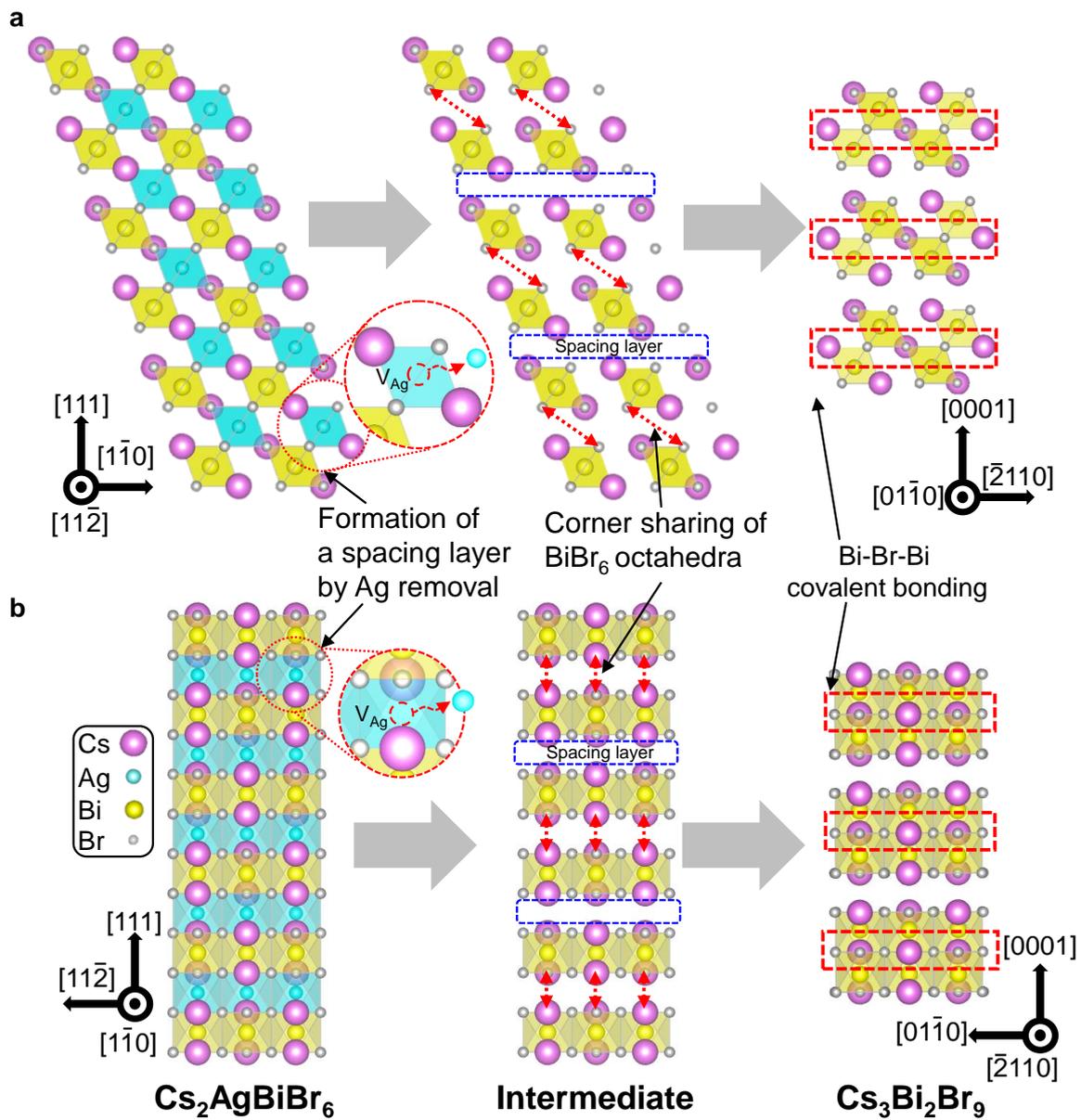

**Figure S14.** Schematic figure for a possible scenario of a structural transition from a double perovskite $Cs_2AgBiBr_6$ phase to a layered perovskite $Cs_3Bi_2Br_9$ phase by Ag-deficiency in $Cs_2AgBiBr_6$ structure.

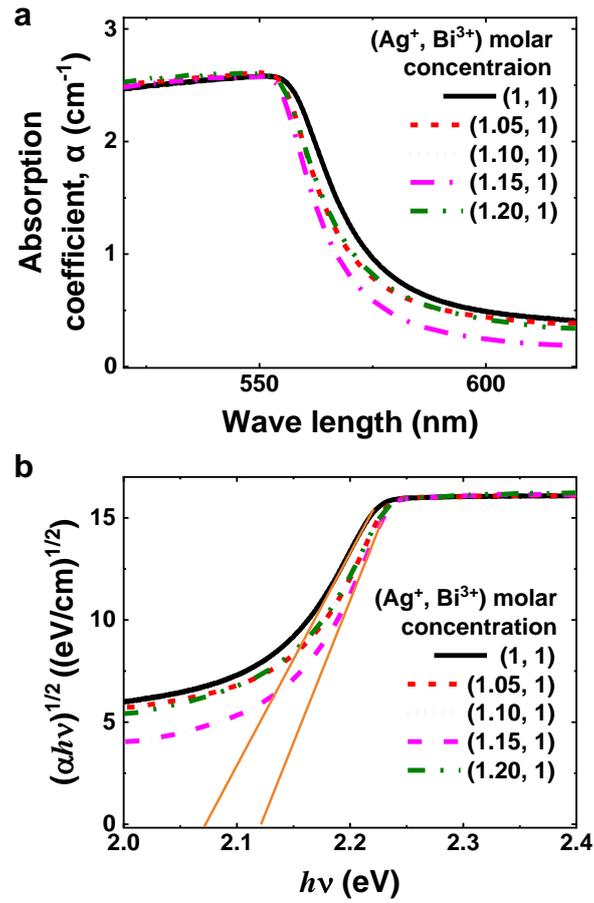

**Figure S15.** (a) The optical absorption coefficient ($\alpha$) of the fundamental absorption edge for the $Cs_2AgBiBr_6$ single crystals grown in stoichiometric and Ag-excess chemical conditions. (b) The absorption coefficient $(\alpha h\nu)^{1/2}$ of the single crystals, and $E_g$ can be determined by extrapolating the linear part to $(\alpha h\nu)^{1/2} = 0$.

**Table S1.** Crystal sizes of $Cs_2AgBiBr_6$ crystals grown under various chemical environments.

| No. of samples | Crystal size (mm) | | | | | | | |
|---|---|---|---|---|---|---|---|---|
| | $(Ag^+, Bi^{3+})$ molar concentration | | | | | | | |
| | (1, 1.20) | (1, 1.15) | (1, 1.10) | (1, 1.05) | (1, 1) | (1.05, 1) | (1.10, 1) | (1.15, 1) | (1.20, 1) |
| **no.1** | 0.4 | 0.5 | 0.6 | 2.0 | 5.2 | 5.1 | 5.2 | 6.6 | 5.8 |
| **no.2** | 0.3 | 0.5 | 0.5 | 1.8 | 5.1 | 4.6 | 5.1 | 6.5 | 5.7 |
| **no.3** | 0.3 | 0.4 | 0.5 | 1.7 | 4.7 | 4.3 | 5.0 | 5.4 | 4.8 |
| **no.4** | 0.3 | 0.4 | 0.4 | 1.5 | 4.5 | 4.2 | 4.8 | 4.6 | 4.6 |
| **no.5** | 0.3 | 0.4 | 0.4 | 1.5 | 4.3 | 4.1 | 4.5 | 4.1 | 3.9 |
| **no.6** | 0.3 | 0.4 | 0.4 | 1.4 | 4.3 | 4.0 | 4.3 | 4.0 | 3.8 |
| **no.7** | 0.3 | 0.3 | 0.4 | 1.4 | 4.2 | 3.9 | 4.2 | 3.9 | 3.8 |
| **no.8** | 0.3 | 0.3 | 0.4 | 1.4 | 3.7 | 3.9 | 4.2 | 3.8 | 3.7 |
| **no.9** | 0.2 | 0.3 | 0.4 | 1.4 | 3.6 | 3.9 | 3.8 | 3.7 | 3.6 |
| **no.10** | 0.2 | 0.3 | 0.3 | 1.3 | 3.5 | 3.9 | 3.8 | 3.7 | 3.6 |
| **no.11** | 0.2 | 0.3 | 0.3 | 1.2 | 1.8 | 3.5 | 3.6 | 3.3 | 3.4 |
| **no.12** | 0.2 | 0.2 | 0.3 | 1.1 | 1.5 | 3.2 | 3.5 | 2.8 | 3.2 |
| **no.13** | 0.2 | 0.2 | 0.2 | 0.8 | 1.3 | 3.1 | 2.6 | 2.6 | 3.1 |
| **no.14** | 0.2 | 0.2 | 0.2 | 0.7 | 1.2 | 2.9 | 2.5 | 2.5 | 3.1 |
| **no.15** | 0.2 | 0.2 | 0.2 | 0.6 | 1.1 | 2.8 | 2.5 | 2.2 | 3.1 |
| **AVG** | 0.26 | 0.33 | 0.37 | 1.32 | 3.33 | 3.83 | 3.97 | 3.98 | 3.95 |
| **SD** | 0.06 | 0.10 | 0.11 | 0.38 | 1.46 | 0.61 | 0.88 | 1.29 | 0.86 |
| **SE** | 0.02 | 0.03 | 0.03 | 0.10 | 0.38 | 0.16 | 0.23 | 0.33 | 0.22 |

**Table S2.** Estimated indirect band gap of the $Cs_2AgBiBr_6$ single crystals grown in stoichiometric and Ag-excess chemical conditions by extrapolating the linear part to $(\alpha h\nu)^{1/2} = 0$ from Tauc plot (Figure S15b)

| ($Ag^+$, $Bi^{3+}$) molar ratio | Indirect band gap (eV) |
|---|---|
| (1, 1) | 2.07 |
| (1.05, 1) | 2.10 |
| (1.10, 1) | 2.11 |
| (1.15, 1) | 2.12 |
| (1.20, 1) | 2.10 |

**References**

[1] Tauc J, Grigorovici R, Vancu A. Optical properties and electronic structure of amorphous germanium. Phys Status Solidi B 1966;15:627-637.

[2] Chiu F-C. A review on conduction mechanisms in dielectric films. Adv Mater Sci Eng 2014;2014:578168.

[3] Jain A, Kumar P, Jain S, Kumar V, Kaur R, Mehra R. Trap filled limit voltage ($V_{TFL}$) and $V^2$ law in space charge limited currents. J Appl Phys 2007;102:094505.

[4] Lampert MA. Simplified theory of space-charge-limited currents in an insulator with traps. Phys Rev 1956;103:1648-1656.

[5] Mott NF, Gurney RW. Electronic processes in ionic crystals. J Chem Educ 1940;18:249.

[6] Kresse G, Furthmüller J. Efficiency of ab-initio total energy calculations for metals and semiconductors using a plane-wave basis set. Comput Mater Sci 1996;6:15-50.

[7] Kresse G, Joubert D. From ultrasoft pseudopotentials to the projector augmented-wave method. Phys Rev B 1999;59:1758-1775.

[8] Perdew JP, Wang Y. Accurate and simple analytic representation of the electron-gas correlation energy. Phys Rev B 1992;45:13244-13249.

[9] Kang J, Tongay S, Zhou J, Li J, Wu J. Band offsets and heterostructures of two-dimensional semiconductors. Appl Phys Lett 2013;102:012111.

[10] Heyd J, Scuseria GE, Ernzerhof M. Hybrid functionals based on a screened Coulomb potential. J Chem Phys 2003;118:8207-8215.

[11] McClure ET, Ball MR, Windl W, Woodward PM. $Cs_2AgBiX_6$ ($X$= Br, Cl): new visible light absorbing, lead-free halide perovskite semiconductors. Chem Mater 2016;28:1348-1354.

[12] Volonakis G, Haghighirad AA, Milot RL, Sio WH, Filip MR, Wenger B, Johnston MB, Herz LM, Snaith HJ, Giustino F. $Cs_2InAgCl_6$: a new lead-free halide double perovskite with direct band gap. J Phys Chem Lett 2017;8:772-778.

[13] Savory CN, Walsh A, Scanlon DO. Can Pb-free halide double perovskites support high-efficiency solar cells? ACS Energy Lett 2016;1:949-955.

[14] Sze SM, Ng KK. Physics of Semiconductor Devices. New Jersy: John Wiley & Sons; 2007.